\documentclass[aps,prl,twocolumn,showpacs,superscriptaddress]{revtex4-1}
\usepackage{amsmath}
\usepackage{epsfig}
\usepackage{color}
\usepackage{endnotes}
\usepackage{natbib}
\usepackage[colorlinks,breaklinks]{hyperref}
\usepackage{nicefrac}
\usepackage{bm}
\usepackage{dcolumn}
\usepackage{graphics}
\usepackage{graphicx} 
\usepackage{tikz} 
\usepackage{amsmath}
\usepackage{amssymb}
\usepackage{color}
\usepackage{changes}
\usepackage{multirow}
\usepackage[nowatermark]{fixmetodonotes}

\usepackage{titlesec}
\titlespacing*{\section}{0pt}{0.8\baselineskip}{0.8\baselineskip}

\begin{document}
\def\Label{}

\title{Inclusive Electron Scattering And The GENIE Neutrino Event Generator}

\newcommand*{\MIT }{Massachusetts Institute of Technology, Cambridge, Massachusetts 02139, USA}
\newcommand*{\MITindex}{1}
\affiliation{\MIT}
\newcommand*{\FNAL}{Fermi National Accelerator Laboratory, Batavia, IL}
\newcommand*{\FNALindex}{2}
\affiliation{\FNAL}
\newcommand*{\PITT}{University of Pittsburgh, Pittsburgh, PA}
\newcommand*{\PITTindex}{3}
\affiliation{\PITT}
\newcommand*{\ODU}{Old Dominion University, Norfolk, Virginia 23529}
\newcommand*{\ODUindex}{4}
\affiliation{\ODU}
\newcommand*{\TAU }{School of Physics and Astronomy, Tel Aviv University, 
Tel Aviv 69978, Israel}
\newcommand*{\TAUindex}{5}
\affiliation{\TAU}
\newcommand*{\CERN}{CERN, European Organization for Nuclear Research, Geneva, Switzerland}
\newcommand*{\CERNindex}{6}
\affiliation{\CERN}
\newcommand*{\UTokyo}{Research Center for Cosmic Neutrinos, Institute for 
Cosmic Ray Research,
  University of Tokyo, Kashiwa, Chiba 277-8582, Japan}
\newcommand*{\UTokyoindex}{7}
\affiliation{\UTokyo}
\newcommand*{\USeville}{Departamento de Fisica Atomica, Molecular y Nuclear, Universidad de Sevilla, 41080 Sevilla, Spain}
\newcommand*{\USevilleindex}{8}
\affiliation{\USeville}

\author{A. Papadopoulou}
\affiliation{\MIT}
\author{A.~Ashkenazi}
\email[Contact Author \ ]{adishka@mit.edu}
\affiliation{\MIT}
\author{S.~Gardiner}
\affiliation{\FNAL}
\author{M.~Betancourt}
\affiliation{\FNAL}
\author{S.~Dytman}
\affiliation{\PITT}
\author{L.B.~Weinstein}
\affiliation{\ODU}
\author{E.~Piasetzky}
\affiliation{\TAU}
\author{F.~Hauenstein}
\affiliation{\ODU}
\affiliation{\MIT}
\author{M. Khachatryan}
\affiliation{\ODU}
\author{S.~Dolan}
\affiliation{\CERN}
\author{G.D.~Megias}
\affiliation{\UTokyo}
\affiliation{\USeville}
\author{O.~Hen}
\affiliation{\MIT}

\date{\today}

\begin{abstract}
\begin{centering}
---------------------
\begin{figure}[!h!]
\centering
\includegraphics[width=0.2\textwidth, height=0.6cm]{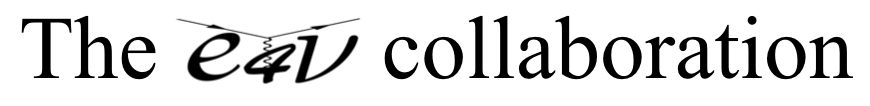}
\end{figure}

\end{centering}
The extraction of neutrino mixing parameters from accelerator-based neutrino
oscillation experiments relies on proper modeling of neutrino-nucleus
scattering processes using neutrino-interaction event generators.  Experimental
tests of these generators are difficult due to
the broad range of neutrino energies produced in accelerator-based beams and
the low statistics of current experiments.  Here we overcome these difficulties by
exploiting the similarity of neutrino and electron interactions with nuclei to
test neutrino event generators using high-precision inclusive electron
scattering data.  To this end, we revised the electron-scattering mode of 
the
GENIE event generator ($e$-GENIE) to include electron-nucleus bremsstrahlung
radiation effects and to use, when relevant, the exact same physics models and
model parameters, as the standard neutrino-scattering version.  We also
implemented new models for quasielastic (QE) scattering and meson exchange
currents (MEC) based on the theory-inspired SuSAv2 approach. Comparing the new $e$-GENIE
predictions with inclusive electron scattering data, we find an overall
adequate description of the data in the QE- and MEC-dominated lower energy
transfer regime, especially when using the SuSAv2 models. Higher energy
transfer-interactions, which are dominated by resonance production, are still
not well modeled by $e$-GENIE.
\end{abstract}

\maketitle



\newpage
\section{Introduction}

The extraction of neutrino mixing parameters from neutrino oscillation
experiments~\cite{nova19,T2K,DUNE} relies on comparing the energy-dependent 
neutrino event
distribution for a particular neutrino flavor near the neutrino production point 
with that at a significant distance away. In practice, the yield at
each neutrino energy is extracted from the measured neutrino-nucleus 
interactions in a detector, as reconstructed from the measured particles 
ejected in the neutrino-nucleus interaction.  This requires detailed knowledge 
of the $\nu$-nucleus interaction.

\begin{figure} [ht]
  \centering
\includegraphics[width=0.48\textwidth]{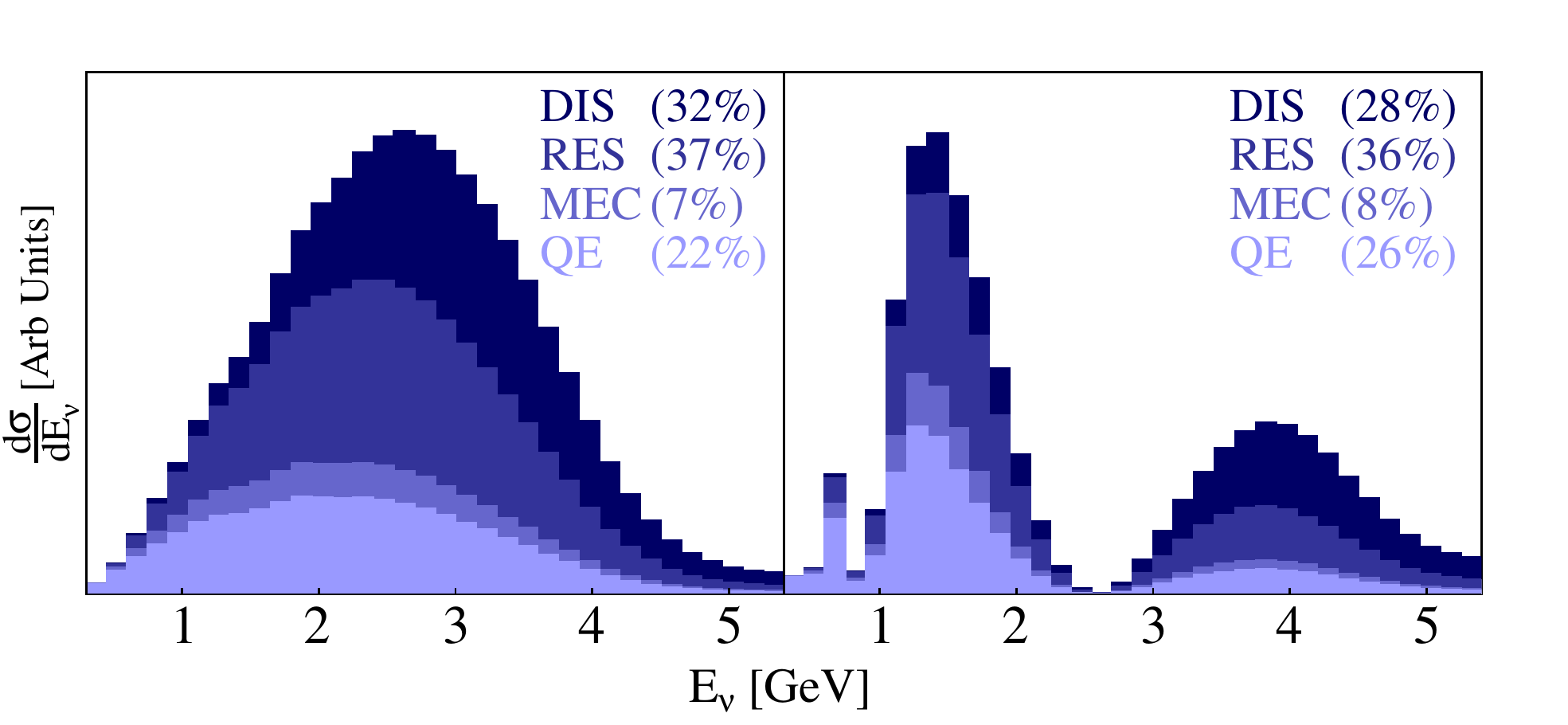}
\caption{\label{fig:Fluxes} Charged-current cross sections as a function
  of neutrino energy obtained using GENIE for muon neutrino scattering
  using the DUNE near detector (left) and far detector (right) oscillated
  fluxes~\cite{DUNEFlux}.  The shaded bands show the
  fractional contribution for each interaction mechanism,
  quasi-elastic scattering (QE), meson-exchange currents (MEC),
  resonance excitation (RES), and deep inelastic scattering (DIS).
  See text for details of the interaction mechanisms. The numbers in
  parentheses indicate the percentage of the cross section due to each
  interaction mechanism.}
\end{figure}

Unfortunately, measuring
 the $\nu$-nucleus interaction is difficult due to the wide energy
 spread of accelerator-produced neutrino beams (see, e.g.,
 Fig.~\ref{fig:Fluxes}(left)) and the tiny
 $\nu$-nucleus cross section.  A relatively small body of data has been 
published~\cite{Alvarez-Ruso:2017oui}, which suffers from poor statistics
and is flux-averaged over a wide range of neutrino energies. This data is 
then
supplemented with theoretical models and implemented into event generator 
codes such as GENIE~\cite{Genie2010}
to simulate the $\nu$-nucleus interaction across a wide range of energies 
and targets.  GENIE simulations are then used to aid in extraction of the 
incident neutrino flux
as a function of energy from the $\nu$-nucleus scattering events
measured in neutrino detectors.

However, the theoretical models need to describe many different interaction
processes for medium to heavy nuclei (typically C, O, or Ar) where 
nuclear effects complicate the interactions.  As a result, the
uncertainties in the extraction of oscillation parameters are often
dominated by lack of knowledge of the $\nu$-nucleus
interaction~\cite{T2K,nova19}.

Fig. \ref{fig:Fluxes} shows such a wide energy spectrum for the DUNE near
detector flux-averaged cross sections (left) and the far detector
oscillated flux-averaged cross sections (right) using one
model configuration in GENIE.  All four 
$\nu$-nucleus reaction mechanisms contribute significantly and all
four need to be well understood.  This is especially true because
different reaction mechanisms contribute differently in the different
oscillation peaks.  Understanding one reaction mechanism better than
the others could have significant oscillation-analysis implications.

\begin{figure}[b]
\centering
\includegraphics[width=0.3\textwidth]{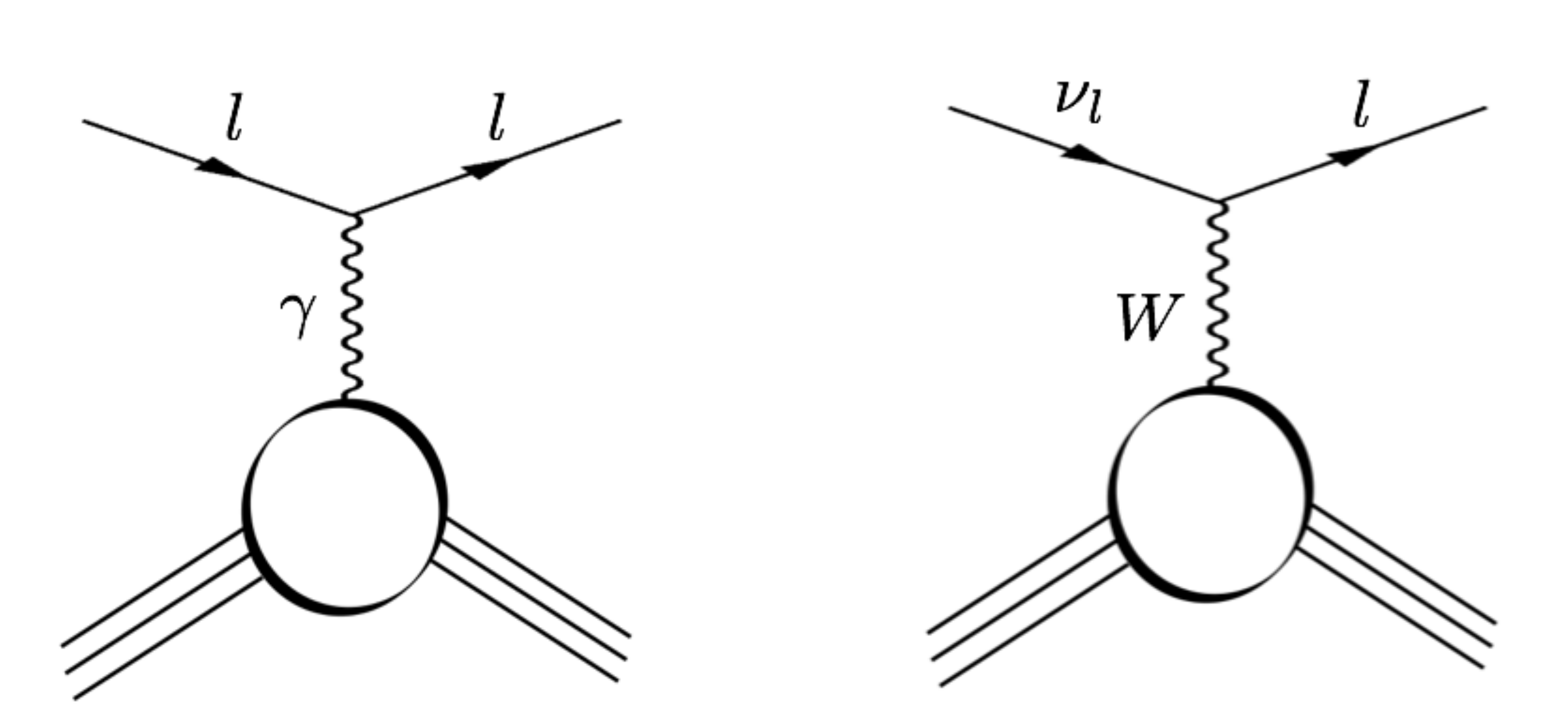}
\caption{(left) electron-nucleus inclusive scattering via one-photon
  exchange and (right) charged current neutrino-nucleus inclusive
  scattering via $W$ exchange with a final state charged lepton.}
\label{fig:eAnuA}
\end{figure}

\begin{figure}[htpb]
\centering
\includegraphics[height=2cm, width=0.49\textwidth]{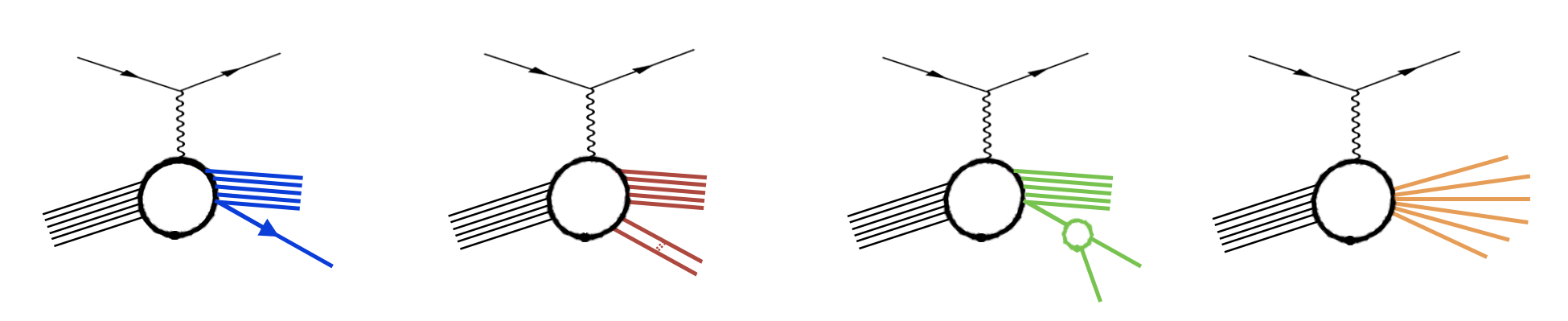}
\caption{Reaction mechanisms for lepton-nucleus scattering (a)
  quasielastic scattering (QE) where one nucleon is knocked out of the
  nucleus, (b) $2p2h$ where two nucleons are knocked out of the
  nucleus, (c) RES resonance production where a nucleon is excited to
  a resonance which decays to a nucleon plus meson(s), and (d) DIS where
 the lepton interacts with a quark in the nucleon.}
\label{fig:ReacMech}
\end{figure}

Because neutrinos and electrons are both leptons, they interact with
atomic nuclei in similar ways (see Fig.~\ref{fig:eAnuA}).  Electrons
interact via a vector current 
($j^\mu_{EM}=\bar u\gamma^\mu u$) 
and neutrinos interact via vector and axial-vector 
($j_{CC}^\mu=\bar u\gamma^\mu(1 - \gamma^5)u\frac{-ig_W}{2\sqrt{2}}$) 
currents.

This gives an inclusive $(e,e')$ electron-nucleon scattering
cross section that depends on only two structure functions:
\begin{eqnarray}
\frac{d^2\sigma^{e}}{dxdQ^2}=\frac{4\pi\alpha^2}{Q^4}\left[\frac{1-y}{x}F^{e}_2(x,Q^2)
  + y^2F^{e}_1(x,Q^2)\right] \,.
\label{eq:sigmaeA}
\end{eqnarray}
Here $F^{e}_1$ and $F^{e}_2$ are the standard electromagnetic vector
structure functions, $Q^2={\bf q}^2-\omega^2$ is the squared momentum
transfer and {\bf q} and $\omega$ are the three-momentum and energy
transfers, $x=Q^2/(2m\omega)$ is the Bjorken scaling variable, $m$ is the
nucleon mass, $y=\omega/E_e$ is the electron fractional energy loss, and
$\alpha$ is the fine structure constant.  This formula is valid for
$Q^2\gg m^2$ where the electron-nucleon cross section is simplest.
Cross sections at lower $Q^2$ have more complicated factors
multiplying each of the two structure functions.

The corresponding inclusive charged current (CC) $(\nu,l^\pm)$ neutrino-nucleon
cross section (where $l^\pm$ is the outgoing charged lepton) has a
similar form with the addition of third, axial, structure function:
\begin{eqnarray}
\begin{split}
\frac{d^2\sigma^{\nu}}{dxdQ^2}=&\frac{G_F^2}{2\pi}\left[\frac{1-y}{x}F^{\nu}_2(x,Q^2)
  + y^2F^{\nu}_1(x,Q^2) \right. \\
  &\left. - y(1-y/2)F^{\nu}_3(x,Q^2)\right] \,.
\label{eq:sigmaNuA}
\end{split}
\end{eqnarray}
Here $F^{\nu}_1$ and $F^{\nu}_2$ are parity conserving structure
functions, $F^{\nu}_3$ is a new parity-violating structure function, and
$G_F$ is the Fermi constant.  The parity-conserving structure
functions, $F^{\nu}_1$ and $F^{\nu}_2$, both include a vector-vector
term almost identical to $F^{e}_1$ and $F^{e}_2$ (the electron terms
have both isoscalar and isovector components, but the neutrino terms
have only isovector components), and an additional
axial-axial term.  See
Refs.~\cite{Alvarez-Ruso:2017oui,Katori:2016yel,Amaro:2019zos} for more
detail.

These simple equations are very similar for lepton-nucleus scattering.
In the limit of electron-nucleon elastic scattering ($x=1$), the two
structure functions reduce to the Dirac and Pauli form factors (which
are linear combinations of the electric and magnetic form factors,
$G_E(Q^2)$ and $G_M(Q^2)$).  Neutrino-nucleon elastic scattering has
an additional axial form factor.  In the simplest case where a lepton
scatters quasielastically (QE) from a nucleon in the nucleus and the
nucleon does not reinteract as it leaves the nucleus, then the
lepton-nucleus cross section is the integral over all initial state
nucleons:
\begin{eqnarray}
\begin{split}
\frac{d\sigma}{dE d\Omega}=&\int_{\mathbf{p_i}}\int_{E_b}
d^3\mathbf{p_i} dE_b K
S(\mathbf{p_i},E_b)\frac{d\sigma^{free}}{d\Omega} \\
&\quad\delta^3(\mathbf{q}-\mathbf{p_f}-\mathbf{p_r})\delta(\omega
- E_b - T_f - T_r)
\end{split}
\end{eqnarray}
where $\mathbf{p_i}$ and $\mathbf{p_f}$ are the initial and final
momenta of the struck nucleon (in the absence of reinteraction,
$\mathbf{p_f}=\mathbf{q}+\mathbf{p_i}$), $\mathbf{p_r}=-\mathbf{p_i}$ 
is the
momentum of the recoil $A-1$ nucleus, $E_b$ is the nucleon binding
energy, $S(\mathbf{p_i},E_b)$ is the probability of finding a nucleon in the
nucleus with momentum $\mathbf{p_i}$ and binding energy $E_b$,
$T_f$ and $T_r$ are the kinetic energies of the final state
nucleon and $A-1$ system, $d\sigma^{free}/d\Omega$ is the
lepton-bound nucleon elastic cross section, and $K$ is a known
kinematic factor.  

This simple form is complicated by nucleon
reinteraction which changes the overlap integral between the initial
and final states (and thus the cross section), and changes the
momentum and angle of the outgoing nucleon.

Thus, to calculate even the simplest type of lepton-nucleus
interaction, we need to know the momentum and binding energy distribution 
of
all nucleons in the nucleus, how the outgoing
nucleon wave function is distorted by the nucleon-nucleus potential,
and how the outgoing nucleon kinematics is changed by final state
interactions.  The initial nuclear state and the final state
hadron-nucleus interactions will be identical for $e$- and
$\nu$-nucleus interactions.

In addition, the lepton can knock out two nucleons simultaneously,
either by interacting with a nucleon belonging to a short range
correlated (SRC) pair \cite{Hen:2016kwk} or by interacting with a pair
of nucleons correlated via meson exchange currents (MEC)~\cite{katori2013meson}. And, of course, 
these two interactions add coherently. The lepton can interact with a nucleon,
exciting it to a resonance (RES), which then deexcites, typically resulting in
the emission
of a nucleon plus mesons. The lepton can also
scatter inelastically from a quark in a nucleon (DIS).  All of these
different reaction mechanisms are very similar for electrons and for
neutrinos.  The outgoing hadrons in all of these interactions will
interact identically with the residual nucleus, whether they are
knocked out by an electron or by a neutrino.

MEC are relatively poorly understood and this contributes
significantly to neutrino oscillation uncertainties.
Ref.~\cite{Simo:2016ikv, Gallmeister:2016dnq} showed that inclusive
neutrino MEC cross sections can be calculated directly from the
structure functions in their electron scattering counterparts if the
interacting system is non-relativistic and if only transverse
response functions (i.e. those which concern the spatial
components of the current transverse to the direction of momentum
transfer) contribute  to the cross section.
This latter assumption is justified for electron MEC
interactions by microscopic studies~\cite{Simo:2016ikv} and electron
scattering data analyses. Its application to neutrino scattering data
via the GiBUU theory framework~\cite{Buss:2011mx} provides favourable
comparisons in 2p2h-enhanced regions~\cite{Dolan:2018sbb}. However,
the same microscopic model applied to neutrino scattering suggests
that, while the transverse component generally remains dominant, the
axial component of the longitudinal response function can become
important (especially for anti-neutrinos), breaking the direct link
between electron and neutrino MEC interactions
\cite{Megias:2016fjk}. Despite these shortcomings, electron scattering
can still provide crucial inputs to modelling neutrino MEC
interactions.


Electron-nucleus scattering is much easier to understand than
$\nu$-nucleus scattering for three reasons:
\begin{itemize}
\item Electron beams have a single, well-known, energy;
\item Electron experiments have far less statistical uncertainty because electron
  beams have higher flux and $e$-nucleus cross sections are far
  higher than their $\nu$ counterparts; and
\item Electron cross sections are purely vector.
\end{itemize}
Therefore we can use $e$-nucleus scattering to constrain models of
$\nu$-nucleus scattering.  Any model which fails to accurately
describe $eA$ (vector-vector) scattering data cannot be used with
confidence to simulate $\nu A$ (vector-vector + axial-axial +
vector-axial) interactions.

GENIE started as a neutrino event generator, like almost all event
generators in neutrino physics.  In recognition of the importance of
electron scattering, it was added as a new option in close conjunction
with the neutrino scattering section.  As much as possible, the
neutrino section references vector and axial contributions
separately and uses the same modeling for vector interactions as the
electron section.  Some models were developed separately for electrons
and others were developed for both applications in tandem.


An earlier electron version of GENIE (v2.12) was tested in
Ref.~\cite{Ankowski:2020qbe} by comparing with inclusive $(e,e')$
data.  Although the quasielastic peak was well-described for a variety
of energies and nuclei, the resonance region (total hadronic energy
$W> 1.1$ GeV) was poorly described.  However, the establishment of
full compatibility between the electron and neutrino versions was then
still in its early stages.

\begin{figure}[t]
\centering
\includegraphics[width=0.49\textwidth]{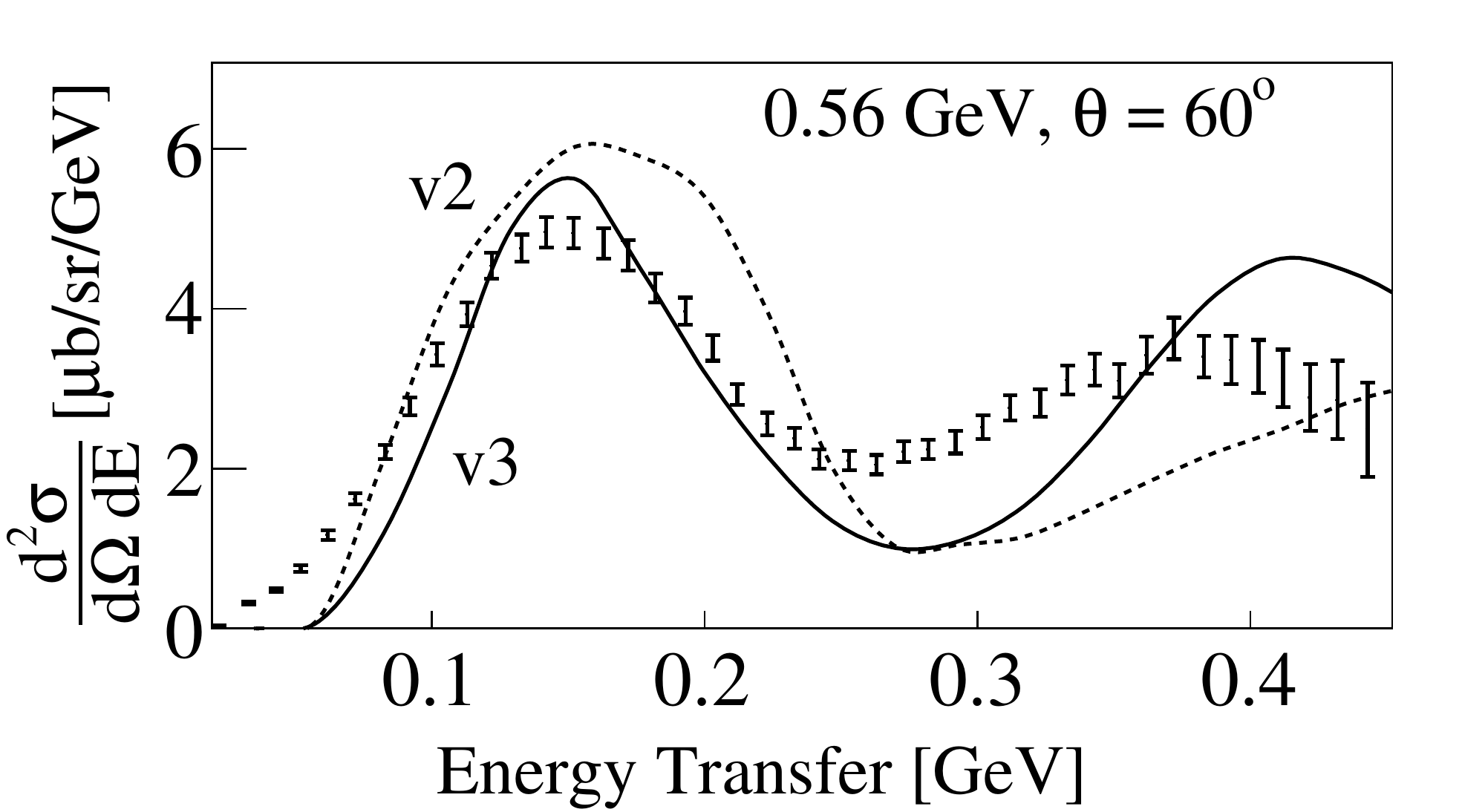}
\caption{Comparison between GENIE v2 and v3 descriptions of inclusive
  C$(e,e')$ scattering cross sections  at $E_0=0.56$ GeV, $\theta_e = 60^\circ$ and $Q^2_{QE}\approx
  0.24$ GeV$^2$~\cite{Barreau:1983ht}.  Black points show the data, solid 
black line 
  shows the GENIE v3 results and dashed black line shows the GENIE
  v2 results.}
\label{fig:Improvements}
\end{figure}

Here we significantly improved both neutrino and electron versions of
GENIE to address these and other issues.  We fixed significant errors
in the previous version, including an error in the Mott cross section
in the electron QE Rosenbluth interaction, a missing Lorentz boost in
the MEC interaction (for both $e$ and $\nu$ interactions), and
incorrect electron couplings used in the RES interactions.  
We worked to better integrate the electron and neutrino codes for quasielastic
and 2p2h models.
We also added more up-to-date models such as
SuSAv2~\cite{Amaro:2019zos}.  These changes have been incorporated in
the latest GENIE version.  We refer to the electron-scattering
component of the widely-used GENIE~\cite{Genie2010} event generator as
$e$-GENIE.

The GENIE improvements
can be seen in Fig.~\ref{fig:Improvements}.  The older v2 QE peak (at
$\omega\approx 0.15$) is too large and is slightly shifted to higher
energy transfer than the data and the first resonance peak is at much
too large an energy transfer.  The updated v3 QE peak has about the
correct integral and is at the correct energy transfer (but is
slightly too narrow) and the first resonance peak is located at
$m_\Delta - m \approx 300$ MeV beyond the QE peak, as expected.
Details of the calculations and of the discrepancies between GENIE v3 and the
data are discussed in detail below.

We specifically focus on testing our knowledge of the electron-nucleus cross
section by benchmarking $e$-GENIE against existing inclusive electron
scattering data for different target nuclei, beam energies and
scattering angles.  The goals are very similar to
Ref.~\cite{Ankowski:2020qbe}, but we test a much more modern version
of $e$-GENIE and we also compare different models within $e$-GENIE.
In addition, if $e$-GENIE describes electron-nucleus scattering well, then it
would be an improvement on the former empirical
fit~\cite{Bosted:2012qc} and would be valuable for helping simulate a
variety of electron experiments.


\section{Modeling} 
\label{sec:modeling}

The most common lepton-nucleus interaction mechanisms include
(Fig.~\ref{fig:ReacMech}): (a) quasielastic (QE) scattering from
individual moving nucleons in the nucleus; (b) two-nucleon knockout,
due to interactions with a
meson being exchanged between two nucleons 
(referred to two-particle two-hole excitations,
$2p2h$ or its major component, meson exchange currents, MEC); 
(c) interactions which leave the struck nucleon in an
excited state (resonance production or RES); and (d) non-resonant
interactions with a quark within the nucleon (DIS).

However, GENIE does not include interference between the amplitudes of different
reaction modes, i.e., the total cross section is obtained by adding the
individual cross sections $\sigma_i(E)$ incoherently.  

For fixed incident beam energy and scattered electron angle, the dominant
process changes from QE at low energy transfer ($\omega\approx Q^2/2m$)
through MEC to RES and to DIS at high energy transfer. Therefore, examining the
agreement of $e$-GENIE with data as a function of energy transfer can provide
valuable insight into the specific shortcomings of the $e$-GENIE models and
their implementations.  This separation according to the underlying physics 
interactions gives valuable insights which are not presently possible with
neutrino cross sections, because only broad-energy beams are available.

The GENIE simulation framework offers several models of the nuclear ground
state, several models for each of the $eA$ or $\nu A$ scattering mechanisms
(each with various tunable model parameters), and several models for hadronic
final state interactions (FSI), i.e., intranuclear rescattering of the outgoing
hadrons~\cite{Dytman:2021ohr,Genie2010,Andreopoulos:2015wxa}.  In this section, we describe
the different models relevant for this work and the electron-specific effects
that we accounted for during $e$-GENIE development.

Since our goal is to use electron scattering data to validate neutrino
interaction modeling in GENIE, the GENIE code for
electron and neutrino interactions are unified in many places.
 The neutrino
interacts with a nucleus via the weak interaction and massive $W$ or
$Z$ exchange, whereas the electron interacts mostly
electromagnetically via massless photon exchange, see
Fig.~\ref{fig:eAnuA}.   This causes the
 cross sections 
to differ by an overall factor of
\begin{equation}
  \frac{8\pi^2\alpha^2}{G_F^2}\frac{1}{Q^4}
  \label{eq:factor}
\end{equation}
(see Eqs.~\ref{eq:sigmaeA} and \ref{eq:sigmaNuA}).  In the code, both
interactions use the same nuclear ground state and many of the nuclear
reaction effects (e.g. FSI) are very similar or identical.  identical.
Except for mass effects and form factors, the electron nucleus cross
section can be obtained by setting the axial part of the interaction
to zero.  We also accounted for isoscalar and isovector terms
appropriately.



Many of the models reported in this work (except for SuSAv2) use the GENIE
implementation of the local Fermi gas (LFG) model to describe the
nuclear ground state. In the simplest Fermi gas model, nucleons occupy
all momentum states up to the global Fermi momentum $k_F$ with equal
probability.  In the LFG model, the Fermi momentum at a given radial
position depends on the local nuclear density (obtained from
measurements of nuclear charge densities). To account for this radial
dependence, GENIE selects an initial momentum for the struck nucleon
by first sampling an interaction location $r$ inside the nucleus
according to the nuclear density. The nucleon momentum is then drawn
from a Fermi distribution using the local Fermi momentum $k_F(r)$.

Another commonly used nuclear model is the Relativistic Fermi Gas
(RFG). Here a global momentum distribution is used for the entire
nucleus, independent of the interaction location in the
nucleus. However, a high-momentum tail of nucleons with momenta above
the Fermi-momentum is included.  This tail is meant to approximately
account for the effects of two-nucleon short-range
correlations~\cite{hen15b, Hen:2016kwk} and follows a $1/k^4$ distribution, where
$k$ is the nucleon momentum.

We consider two distinct sets of GENIE models for QE and MEC: 
\begin{itemize}
\item {\bf G2018}, which uses the Rosenbluth cross section with the
  Local Fermi Gas for QE scattering and the Empirical MEC
  model~\cite{Katori:2013eoa}. 
  This model set is formally marked as the \texttt{G18\_10a\_02\_11a}
  configuration of GENIE v3.

\item {\bf GSuSAv2}~\cite{PhysRevD.101.033003}, which follows 
the universal SuSAv2 super-scaling approach to lepton scattering.
This new model set will be included in the forthcoming GENIE v3.2.0 release as the
\texttt{GTEST19\_10b\_00\_000} configuration. 
\end{itemize}
In both model sets, RES is modeled using the Berger-Sehgal
model~\cite{Berger:2007rq} and DIS reactions are modeled using Bodek
and Yang\cite{Bodek2003}.  The models are described in more detail below.

\subsubsection{Quasi Elastic (QE)}
In QE interactions, a lepton scatters on a single nucleon, removing it from the
spectator $A-1$ nucleus unless final-state interactions lead to reabsorption.

The electron QE interaction in the G2018 configuration of GENIE uses
the Rosenbluth cross section with the vector
structure function parametrization of Ref.~\cite{Bradford:2006yz}.  We
corrected the implementation of this model for $e$-GENIE and modified
the cross section as described above. This electron QE cross section
differs in important ways (notably, the Rosenbluth treatment lacks
medium polarization corrections) from the Valencia CCQE model
\cite{ValenciaModel} used in the G2018 configuration for neutrinos.

A new QE model in GENIE, based on the SuSAv2 approach
\cite{Megias:2016lke,PhysRevD.101.033003,Megias:2016fjk}, uses superscaling to write the inclusive
cross section in terms of a universal function (i.e., independent of
momentum transfer and nucleus). For EM scattering, the scaling
function may be expressed in the form
\begin{eqnarray}
f(\psi')=k_F\frac{\frac{d^2\sigma}{d\Omega_e
    d\nu}}{\sigma_{Mott}(v_LG_L^{ee'}+V_TG_T^{ee'})} \,,
\end{eqnarray}
where $\psi'$ is a dimensionless scaling variable, $k_F$ is the
nuclear Fermi momentum, the denominator is the single-nucleon elastic
cross section, $v_L$ and $v_T$ are known functions of kinematic
variables, and $G_L^{ee'}(q,\omega)$ and $G_T^{ee'}(q,\omega)$ are the
longitudinal and transverse nucleon structure functions (linearly related 
to
$F_1^e$ and $F_2^e$)~\cite{Caballero:2006wi}. 
For $e$-GENIE, we extended the original neutrino implementation
\cite{PhysRevD.101.033003} to the electron case using a consistent physics
treatment.

The original SuSAv2 QE cross section calculations used a Relativistic
Mean Field (RMF) model of the nuclear ground
state~\cite{Gonzalez-Jimenez:2019qhq,Gonzalez-Jimenez:2019ejf}. This
approach includes the effects of the real part of the nucleon-nucleus
potential on the outgoing nucleons which creates a ``distorted''
nucleon momentum distribution.

Although GENIE lacks the option to use an RMF nuclear model directly, we
achieve approximate consistency with the RMF-based results by using a two-step
strategy for QE event generation. First, an energy and scattering angle for the
outgoing lepton are sampled according to the inclusive double-differential
cross section. This cross section is computed by interpolating precomputed
values of the nuclear responses $G_L^{ee'}(q,\omega)$ and $G_L^{ee'}(q,\omega)$ which
are tabulated on a two-dimensional grid in $(q, \omega)$ space. The responses were
obtained using the original RMF-based SuSAv2 calculation.

Second, the outgoing nucleon kinematics are determined by choosing its
initial momentum from an LFG distribution. The default nucleon binding
energy used in GENIE for the LFG model is replaced for SuSAv2 with an
effective value tuned to most closely duplicate the RMF
distribution. The outgoing nucleon kinematics are not needed for the
comparisons to inclusive $(e,e')$ data shown in this work.

We also compared those models for QE
scattering to a model using the Llewellyn-Smith CCQE scattering
prescription \cite{LlewellynSmith:1971uhs} and the RFG. 

\subsubsection{Meson Exchange Current (MEC)}

MEC describes an interaction that results in the ejection of two
nucleons from the nucleus (often referred to as $2p2h$).  It typically
proceeds via lepton interaction with a pion being exchanged between
two nucleons or by interaction with a nucleon in an SRC pair.  MEC is
far less understood than other reaction mechanisms because, unlike
the others, it involves scattering from two nucleons simultaneously.
GENIE has several models for MEC.

The G2018 configuration of $e$-GENIE uses the Empirical 
model~\cite{Katori:2013eoa}, that is useable for both $eA$ and $\nu A$
scattering. It assumes that the MEC peak for inclusive scattering has
a Gaussian distribution in $W$ and is located between the QE and first
RES peaks. Although both versions of the model use the same effective 
form factors, the amplitude of the MEC peak was tuned separately to electron
and neutrino scattering data. This model was developed in the context of
empirically fitting GENIE to MiniBooNE inclusive neutrino scattering
data and is still used for neutral-current interactions.

For charged-current neutrino interactions, GENIE G2018 uses the very
different Valencia $2p2h$ model \cite{ValenciaModel,GENIEValenciaMEC}
instead of the Empirical model.

The SuSAv2 model evaluates the $2p2h$ MEC contributions within an exact RFG-based
microscopic calculation that englobes the $2p2h$ states excited by the
action of meson-exchange currents within a fully relativistic
framework
\cite{DePace:2004cr,Simo:2016ikv,Simo:2014wka,Megias:2016fjk}, and
considers the weak vector and axial components for neutrino-nucleus
interactions in both longitudinal and transverse channels as well as a
complete analysis for electromagnetic reactions.  As in the case for
the SuSAv2 QE model, we extended the original GENIE implementation of
SuSAv2 MEC for neutrinos \cite{PhysRevD.101.033003} to the electron
case for $e$-GENIE.
The SuSAv2 MEC model is available for both $eA$ and $\nu A$ scattering
\cite{Megias:2014qva,Amaro:2017eah,Megias:2016lke}.  

\subsubsection{Resonance (RES) and Deep Inelastic Scattering (DIS)}
\label{sec:resdis}
Resonance production in GENIE is simulated using the Berger-Sehgal
model~\cite{Berger:2007rq}, in which the lepton interacts with a
single moving nucleon and excites it to one of 16 resonances.  The
cross sections are calculated based on the Feynman-Kislinger-Ravndal
(FKR) model~\cite{fkr1971}, without any interferences between them.
Form factors are derived separately for vector and axial probes~\cite{Rein:1980wg} 
but have not been updated to include recent electron scattering results.

The GENIE treatment of deep inelastic scattering used in this work is based on
that of Bodek and Yang~\cite{Bodek2003}. Hadronization is modeled using an
approach which transitions gradually between the AGKY model~\cite{Yang2009} and
the PYTHIA 6 model~\cite{PYTHIA6}. At low values of the hadronic invariant mass
$W$, the Bodek-Yang differential cross section is scaled by tunable parameters
that depend on the multiplicity of hadrons in the
final-state~\cite{Andreopoulos:2015wxa}. 

Integration of RES and DIS contributions is complicated by the need
for a model of
non-resonant meson production. 
There is no definite separation of RES and DIS contributions;  GENIE makes a
sharp cutoff at $W=1.93$ GeV in the latest tune and uses a suppression
factor to enable usage of the Bodek-Yang cross section at low $W$ in
place of a true non-resonant model.
These features were recently retuned by
the GENIE collaboration using measurements of charged-current $\nu_\mu$ and
$\bar{\nu}_\mu$ scattering on deuterium~\cite{TenaVidal2018}. 
The W cutoff and suppression factors apply to both $eA$ and $\nu A$ models.

\subsection{Final State Interactions (FSIs)}

Final state interactions of outgoing with hadrons with the residual
nuclei are calculated in eGENIE using the
INTRANUKE~\cite{Dytman:2021ohr,dytman2011fsi} package and one of two
options. The first, hA, an empirical data-driven method, uses the
cross-section of pions and nucleons with {\it nuclei} as a function of
energy up to 1.2 GeV and the CEM03~\cite{mashnik2006fsi} calculation
for higher energies.  The second, hN, is
a full intra-nuclear cascade calculation of the interactions of pions, kaons, photons, and
nucleons with nuclei.  In the hN model, each
outgoing particle can interact successively with any or all the nucleons it
encounters on its path leaving the nucleus, and any particles created
in those interactions can also subsequently reinteract.  The ability
of the two models to describe hadron-nucleus data is very similar.

The $e$-GENIE G2018 configuration uses the hA FSI model,
while GSuSAv2 uses hN. However, the choice of FSI model
has no effect on the inclusive cross sections considered in the present work.

\begin{figure}[t]
\centering
\includegraphics[width=0.4\textwidth]{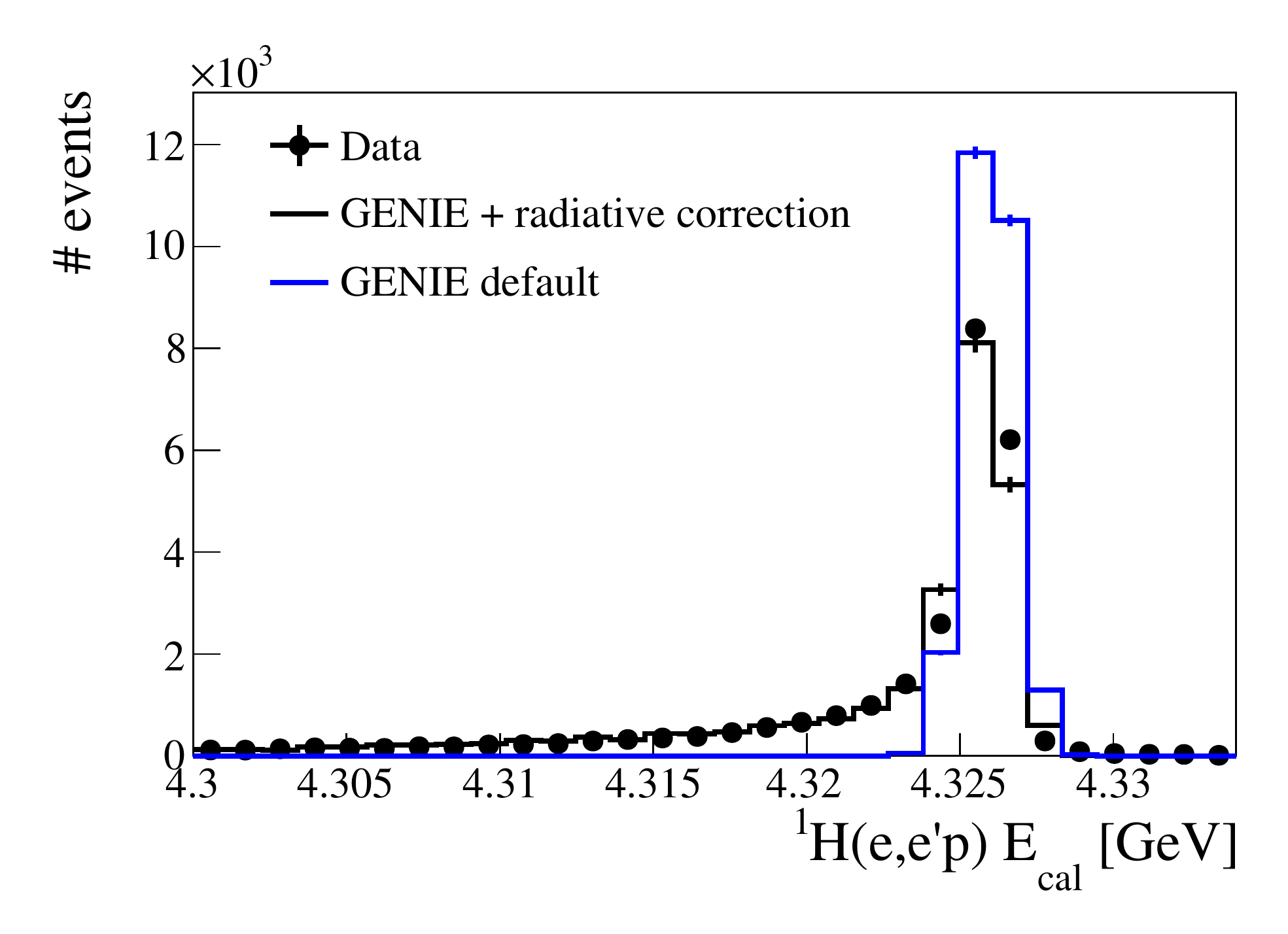}
\caption{  Number of
events vs $E_{cal}= E_{e'} +T_p$ the scattered electron energy plus
proton kinetic energy
  for 4.32 GeV H$(e,e'p)$. Black points are data \cite{Cruz-Torres:2019bqw} , black
  histogram shows the unradiated GENIE prediction and blue histogram
  shows the GENIE prediction with electron radiation.  The GENIE
  calculations have been scaled to have the same integral as the data.}
\label{fig:radTail}
\end{figure}

\begin{figure}[t]
\centering
\includegraphics[width=0.49\textwidth]{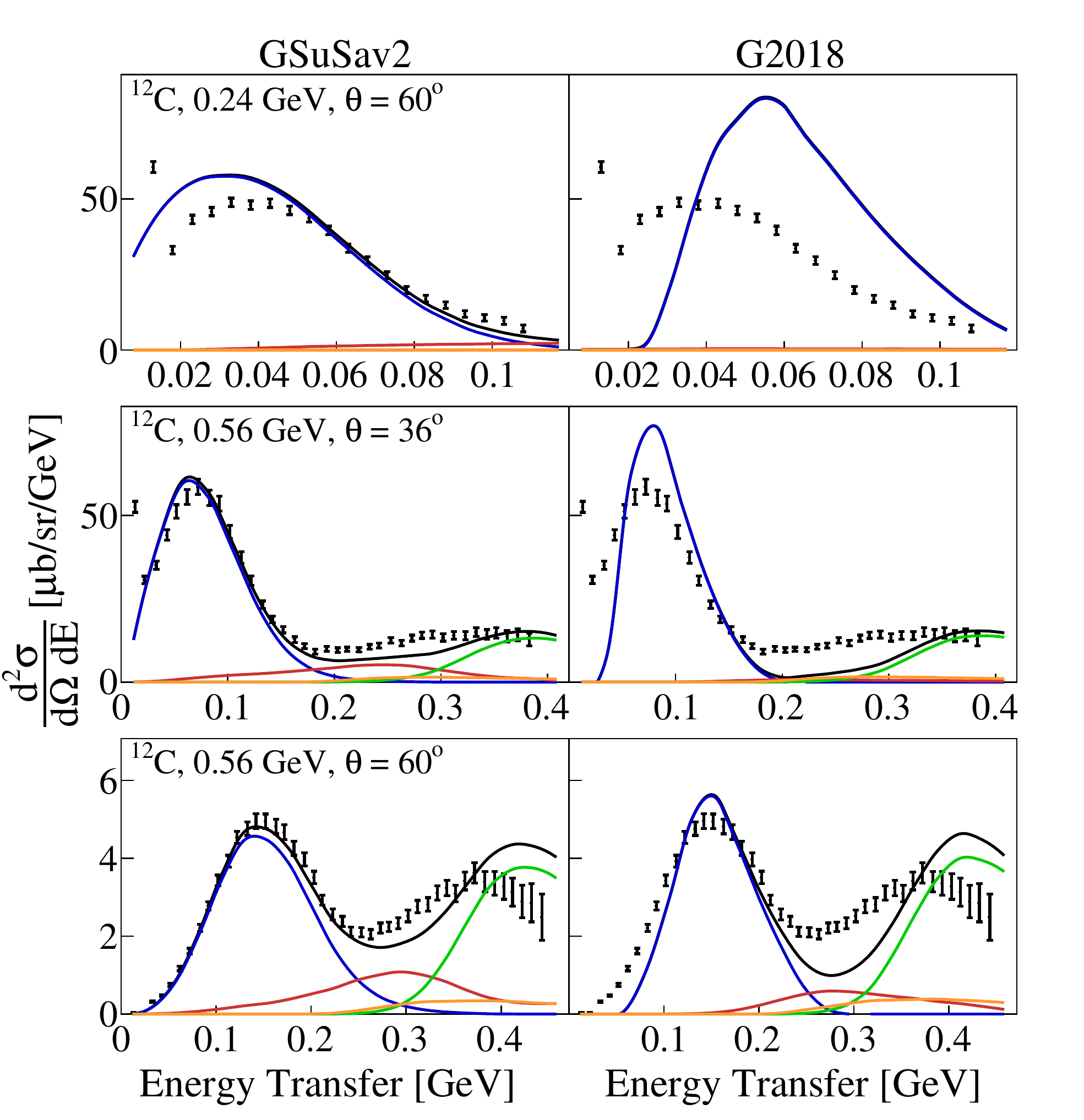}
\caption{Comparison of inclusive C$(e,e')$ scattering cross sections for data
  and for GENIE.  (left) data vs GSuSAv2 and (right) data vs G2018.
  (top) $E_0=0.24$ GeV, $\theta_e = 60^\circ$ and $Q^2_{QE}\approx
  0.05$ GeV$^2$~\cite{Barreau:1983ht}, (middle) $E_0=0.56$ GeV, $\theta_e = 36^\circ$ and $Q^2_{QE}\approx
  0.11$ GeV$^2$~\cite{Barreau:1983ht}, and (bottom) $E_0=0.56$ GeV, $\theta_e = 60^\circ$ and $Q^2_{QE}\approx
  0.24$ GeV$^2$~\cite{Barreau:1983ht}.  Black points show the data, solid 
black lines
  show the total GENIE prediction, colored lines show the contribution
of the different reaction mechanisms: (blue) QE, (red) MEC, (green)
RES and (orange) DIS.}
\label{fig:C1}
\end{figure}

\subsection{Radiative Corrections}

When electrons scatter from nuclei, there are several radiative
effects that change the cross section.  The incoming and outgoing
electrons can each radiate a real photon, which changes the kinematics
of the interaction or the detected particles, and there can be
vertex or propagator corrections that change the cross section.
When comparing electron scattering data to models, either the data or
the model needs to be corrected for radiative effects.  Published
electron scattering cross sections are  typically corrected
for radiative effects, but this correction is complicated and somewhat
model-dependent.

We implemented a framework for electron radiative corrections in GENIE for the
first time (not yet in a GENIE release) to allow comparisons to non-radiatively
corrected data.  The framework allows electron radiation, which can change the
kinematics of the event by changing either the incident or scattered electron
energy (through radiation of a real photon).  We modeled external radiation in
the same way as the Jefferson Lab SIMC event generator \cite{SimcRadiation}. Future
versions of $e$-GENIE will incorporate cross section changes due to vertex and
propagator corrections.

We validated the radiative correction procedure by comparing a
simulated sample to electron scattering from protons at
Jefferson Lab. Figure~\ref{fig:radTail} shows the data
compared to the GENIE simulation with and without radiative
corrections.  The radiatively corrected calculation is clearly much
closer to the data.  The radiative tail of the distribution is only
significant for about 5 MeV below the peak.

This correction can be used for comparisons with non-radiatively-corrected
data.  It was not used to compare with the radiatively-corrected
inclusive data shown below.

\begin{figure}[t]
\centering
\includegraphics[width=0.49\textwidth]{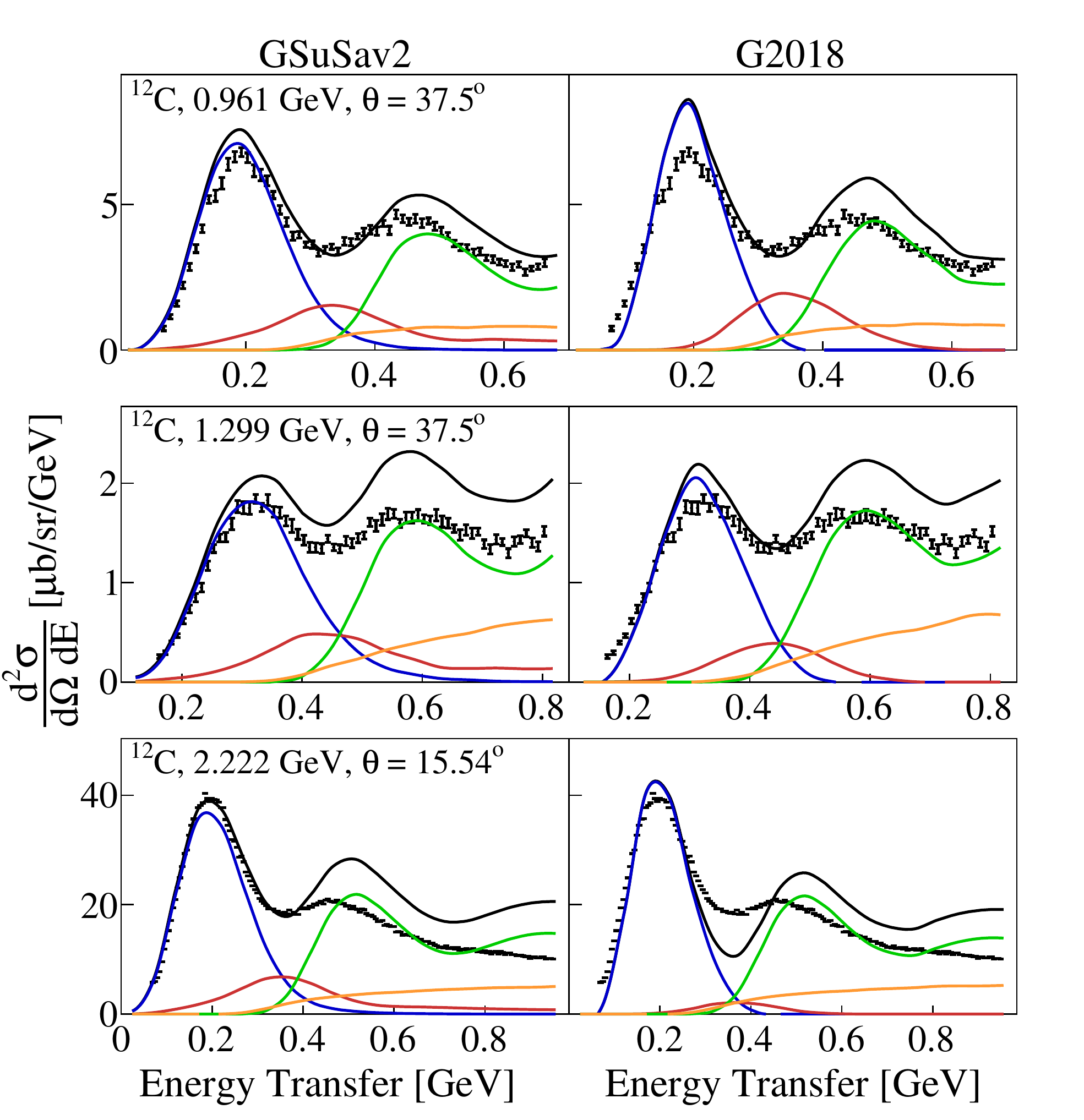}
\caption{Comparison of inclusive C$(e,e')$ scattering cross sections
  for data and for GENIE.  (left) data vs GSuSAv2 and (right) data vs
  G2018.  (top) $E_0=0.96$ GeV, $\theta_e = 37.5^\circ$ and
  $Q^2_{QE}\approx 0.32$ GeV$^2$~\cite{PhysRevLett.62.1350}, (middle)
  $E_0=1.30$ GeV, $\theta_e = 37.5^\circ$ and $Q^2_{QE}\approx 0.54$
  GeV$^2$~\cite{PhysRevLett.62.1350}, and (bottom) $E_0=2.22$ GeV,
  $\theta_e = 15.5^\circ$ and $Q^2_{QE}\approx 0.33$
  GeV$^2$~\cite{Dai:2018xhi}.  Black points show the data, solid black
  lines show the total GENIE prediction, colored lines show the
  contribution of the different reaction mechanisms: (blue) QE, (red) MEC, (green)
RES and (orange) DIS.}
\label{fig:C2}
\end{figure}

\begin{figure}[htbp]
\centering
\includegraphics[width=0.49\textwidth]{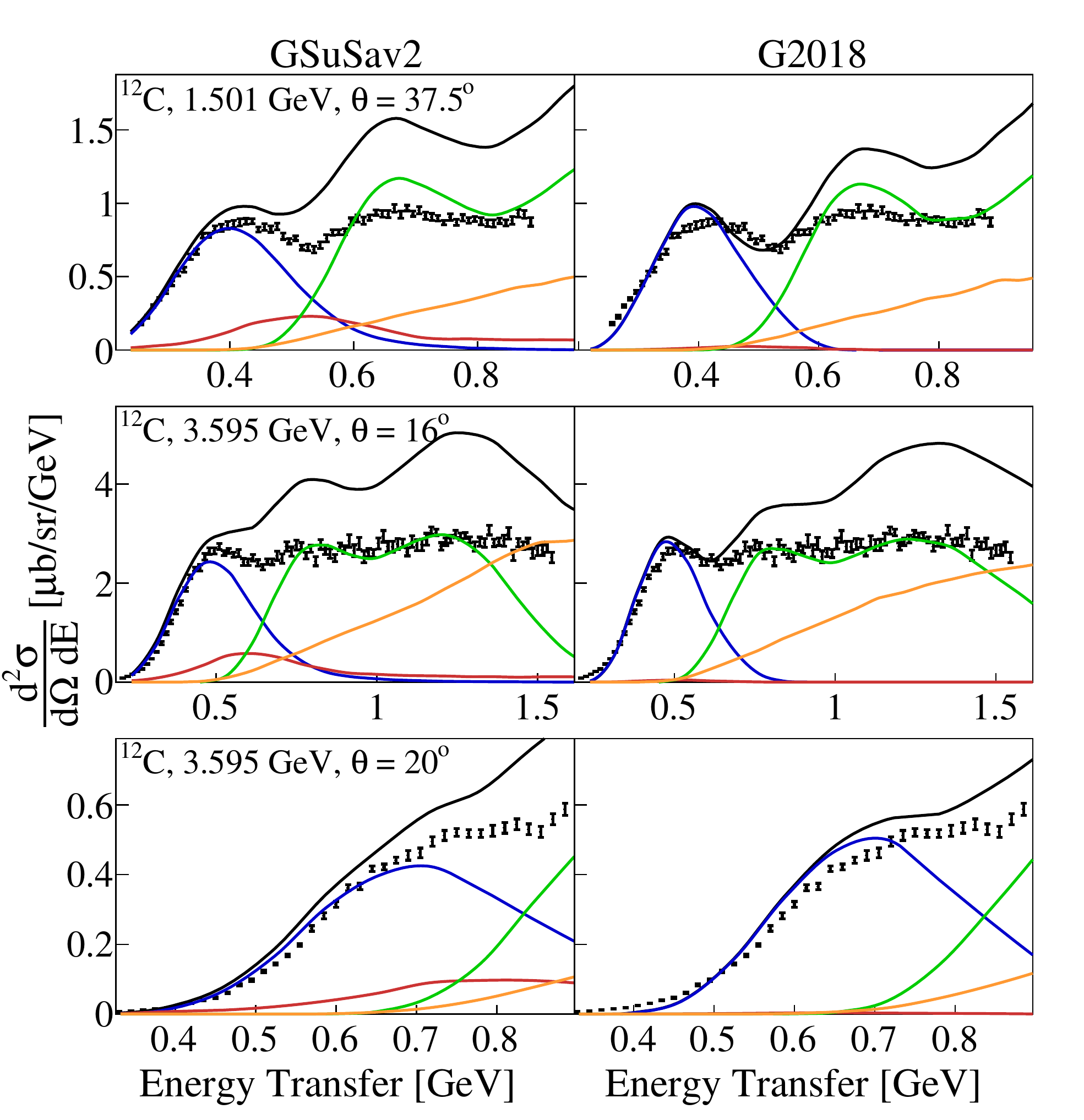}
\caption{Comparison of inclusive C$(e,e')$ scattering cross sections
  for data and for GENIE.  (left) data vs GSuSAv2 and (right) data vs
  G2018.  (top) $E_0=1.501$ GeV, $\theta_e = 37.5^\circ$ and
  $Q^2_{QE}\approx 0.92$ GeV$^2$~\cite{PhysRevLett.62.1350}, (middle)
  $E_0=3.595$ GeV, $\theta_e = 16^\circ$ and $Q^2_{QE}\approx 1.04$
  GeV$^2$~\cite{Day:1993md}, and (bottom) $E_0=3.595$ GeV,
  $\theta_e = 20^\circ$ and $Q^2_{QE}\approx 1.3$
  GeV$^2$~\cite{Day:1993md}.  Black points show the data, solid black
  lines show the total GENIE prediction, colored lines show the
  contribution of the different reaction mechanisms: (blue) QE, (red) MEC, (green)
RES and (orange) DIS.}
\label{fig:C3}
\end{figure}

\section{$e$-GENIE comparisons to inclusive electron scattering data}
\label{sec:comparison}

To test $e$-GENIE, we compare inclusive electron scattering data to theoretical
predictions made using two different program configurations which differ in
their choice of QE and MEC models: G2018 (which adopts the
Rosenbluth model for QE and the empirical Dytman model for MEC) and
GSuSAv2 (which adopts SuSAv2 for both QE and MEC).

Figs.~\ref{fig:C1}, \ref{fig:C2} and \ref{fig:C3} show the inclusive
C$(e,e')$ cross sections for a wide range of beam energies and
scattering angles compared to the G2018 and GSuSAv2 models. The QE
peak is the one at lowest energy transfer ($\nu\approx Q^2/2m$) in
each plot.  The next peak at about 300 MeV larger energy transfer
corresponds to $\Delta(1232)$ excitation and the ``dip region'' is
between the two peaks.  The $\Delta$ peak in the data is separated
from the QE peak by less than the 300 MeV $\Delta$-nucleon mass
difference, indicating that it is shifted in the nuclear medium.

GSuSAv2 clearly describes the QE and dip regions much better
than G2018, especially at the three lowest momentum transfers (see
Fig.~\ref{fig:C1}).  G2018 has particular difficulty describing the
data for $E_0=0.24$ GeV and $\theta_e=60^\circ$, where $Q^2=0.05$
GeV$^2$ at the quasielastic peak.  G2018 also predicts too small a
width for the quasielastic peak and too small a 2p2h/MEC contribution
for $E_0=0.56$ GeV and $\theta_e=60^\circ$; GSuSAv2 describes both
features far better.

\begin{figure}[htbp]
\centering
\includegraphics[width=0.49\textwidth]{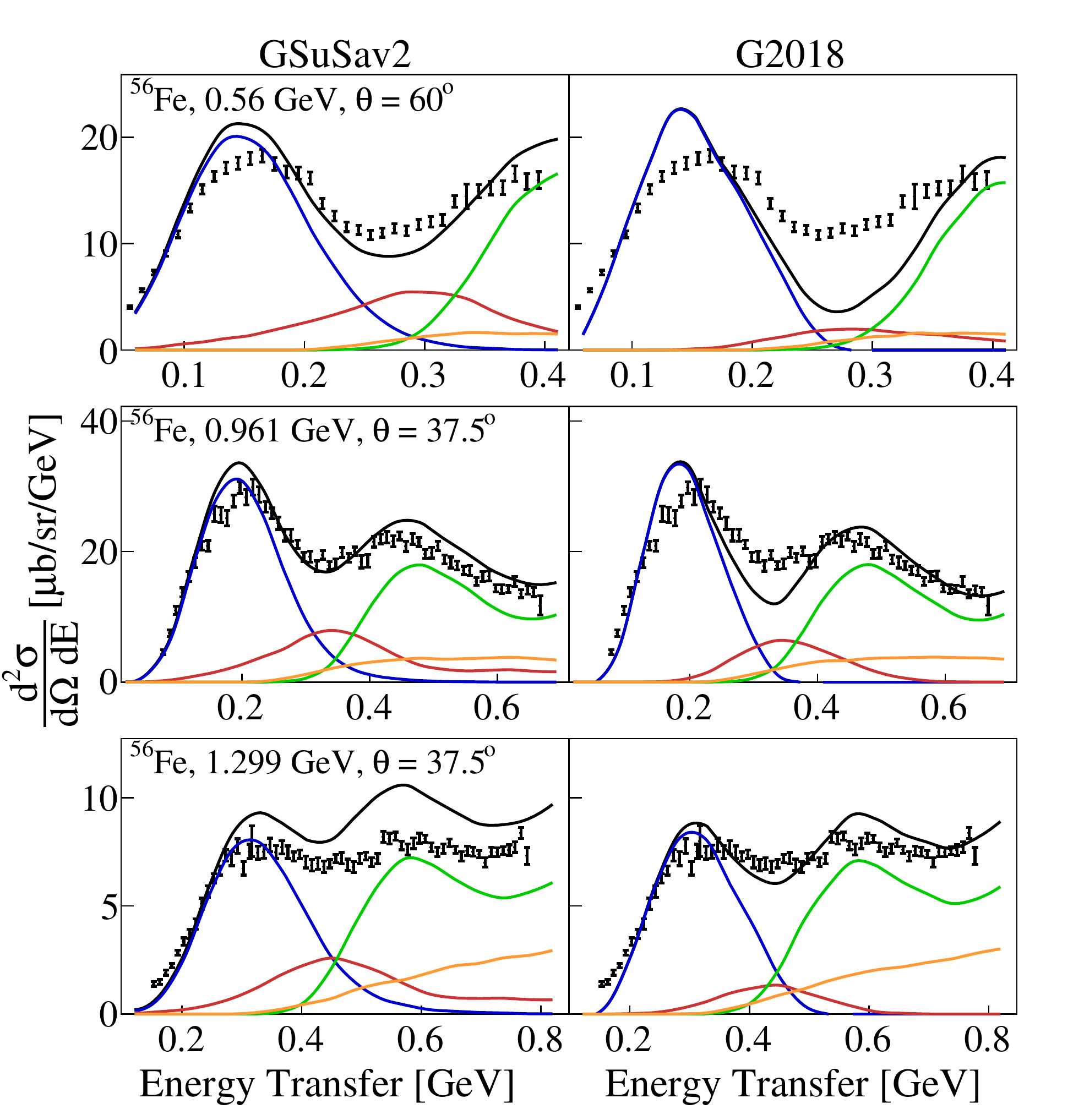}
\caption{Comparison of inclusive Fe$(e,e')$ scattering cross sections
  for data and for GENIE.  (left) data vs GSuSAv2 and (right) data vs
  G2018.  (top) Fe$(e,e')$, $E_0=0.56$ GeV, $\theta_e = 60^\circ$ and
  $Q^2_{QE}\approx 0.24$ GeV$^2$~\cite{Barreau:1983ht}, (middle)
  Fe$(e,e')$, $E_0=0.96$ GeV, $\theta_e = 37.5^\circ$ and
  $Q^2_{QE}\approx 0.32$ GeV$^2$~\cite{PhysRevLett.62.1350}, (bottom)
  Fe$(e,e')$, $E_0=1.30$ GeV, $\theta_e = 37.5^\circ$ and
  $Q^2_{QE}\approx 0.54$ GeV$^2$~\cite{PhysRevLett.62.1350}.  Black
  points show the data, solid black lines show the total GENIE
  prediction, colored lines show the contribution of the different
  reaction mechanisms: (blue) QE, (red) MEC, (green)
RES and (orange) DIS.}
\label{fig:Fe}
\end{figure}

At intermediate momentum transfers (see Fig.~\ref{fig:C2}), GSuSAv2
describes the data somewhat better than G2018, although it overpredicts the dip
region cross section at $E_0=1.299$ GeV and $\theta_e=37.5^\circ$.
The MEC contribution for G2018 appears to be much too small for
$E_0=2.222$ GeV and $\theta_e = 15.54^\circ$ ($Q^2_{QE}=0.33$ GeV$^2$).
Both model sets significantly disagree with the data in the resonance
region (where they use the same RES and DIS models).  The 0.961 GeV,
37.5$^\circ$ and the 2.222 GeV, $15.54^\circ$ data are taken at almost
identical $Q^2_{QE}$.  The lower beam-energy data is more transverse (since
it is at larger scattering angle).  The GSuSAv2 MEC contribution is
similar for both data sets, but the G2018 MEC contribution is far
smaller for the higher beam-energy data.  The GSuSAv2 MEC contribution
describes the dip region better in the higher beam-energy data set.
The RES model appears to agree with the data  slightly better for the
lower beam-energy, more transverse, data set.

At the highest momentum transfers ($Q^2\approx 1$ GeV$^2$), the
disagreement at the larger energy transfers is far greater.  The G2018
``empirical'' MEC model contributions are negligible, in marked contrast to
the GSuSAv2 MEC contributions.  The RES and DIS contributions are 
very significant at high $Q^2$ and in general the GENIE model is larger 
than the data in the region dominated by RES interactions, as noted
in Ref.~\cite{Ankowski:2020qbe}.  In addition, GENIE does not
include the nuclear medium dependent $\Delta$-peak shift, so that the
predicted location of the $\Delta$-peak is at larger energy transfer
than that of the data.

Fig.~\ref{fig:Fe} shows the inclusive Fe$(e,e')$
cross sections for several beam energies and scattering angles
compared to the G2018 and GSuSAv2 models. The GSuSAv2 model describes
the QE region better for all three data sets.  The GSuSAv2 MEC
contributions are significantly larger than the Emprical G2018 MEC
contributions and match the dip-region data far better at $Q^2_{QE}=0.24$ and $0.32$
GeV$^2$.  However, it overpredicts the dip-region cross section at
$Q^2_{QE}=0.54$ GeV$^2$.  The RES and DIS models describe the Fe data
better than the C data at large energy transfers.

\begin{figure}[t]
\centering
\includegraphics[width=0.49\textwidth]{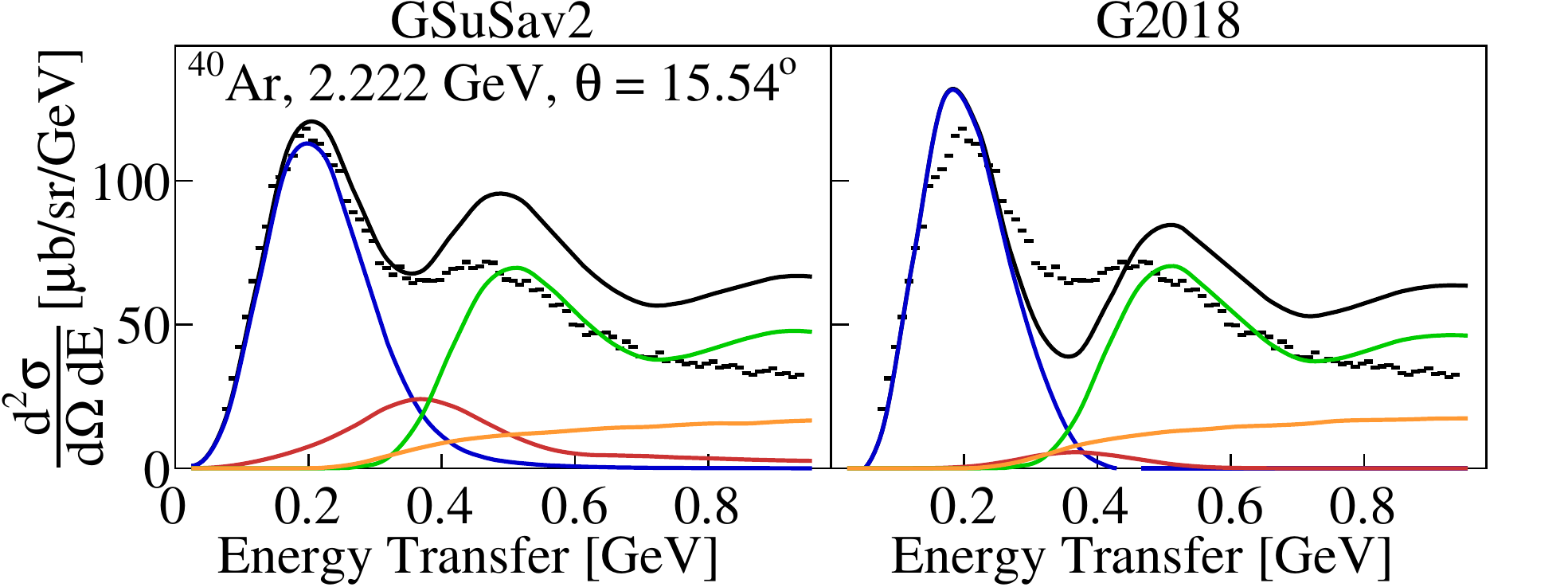}
\caption{Comparison of inclusive Ar$(e,e')$ scattering cross sections for 
data
  and for GENIE at $E_0=2.22$ GeV, $\theta_e = 15.5^\circ$ and $Q^2_{QE}\approx
  0.33$ GeV$^2$~\cite{PhysRevC.99.054608}.  (left) data vs GSuSAv2 and (right) data vs G2018.
   Black points show the data, solid black lines
  show the total GENIE prediction, colored lines show the contribution
of the different reaction mechanisms:  (blue) QE, (red) MEC, (green)
RES and (orange) DIS. }
\label{fig:Ar}
\end{figure}

\begin{figure}[htbp]
\centering
\includegraphics[width=0.49\textwidth]{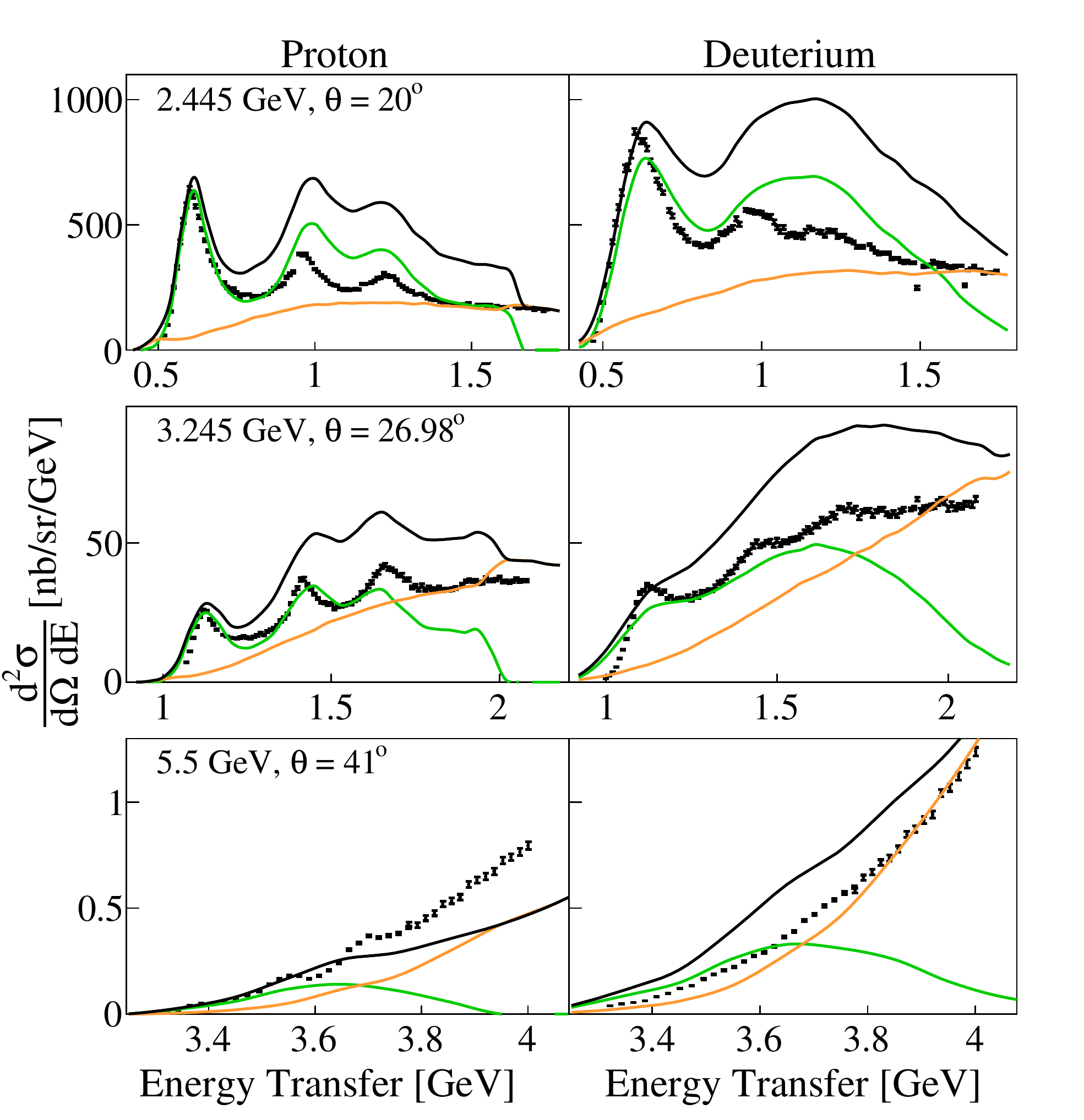}
\caption{Comparison of inclusive proton (left) and deuterium (right)
  $(e,e')$ scattering cross sections for data and for GENIE using
  G2018. (top) $E_0=2.445$ GeV and $\theta_e = 20^\circ$, (middle)
  $E_0=3.245$ GeV and $\theta_e = 26.98^\circ$, and (bottom) $E_0=5.5$
  GeV and
  $\theta_e =
  41^\circ$~\cite{nicolescu2009,PhysRevLett.85.1186,malace2006}.
  Black points show the data, solid black lines show the total GENIE
  prediction, colored lines show the contribution of the different
  reaction mechanisms: (green)
RES and (orange) DIS.  The first peak at lowest energy transfer is the
$\Delta(1232)$ resonance.}
\label{fig:Hydrogen}
\end{figure}

\begin{figure}[htbp]
\centering
\includegraphics[width=0.49\textwidth]{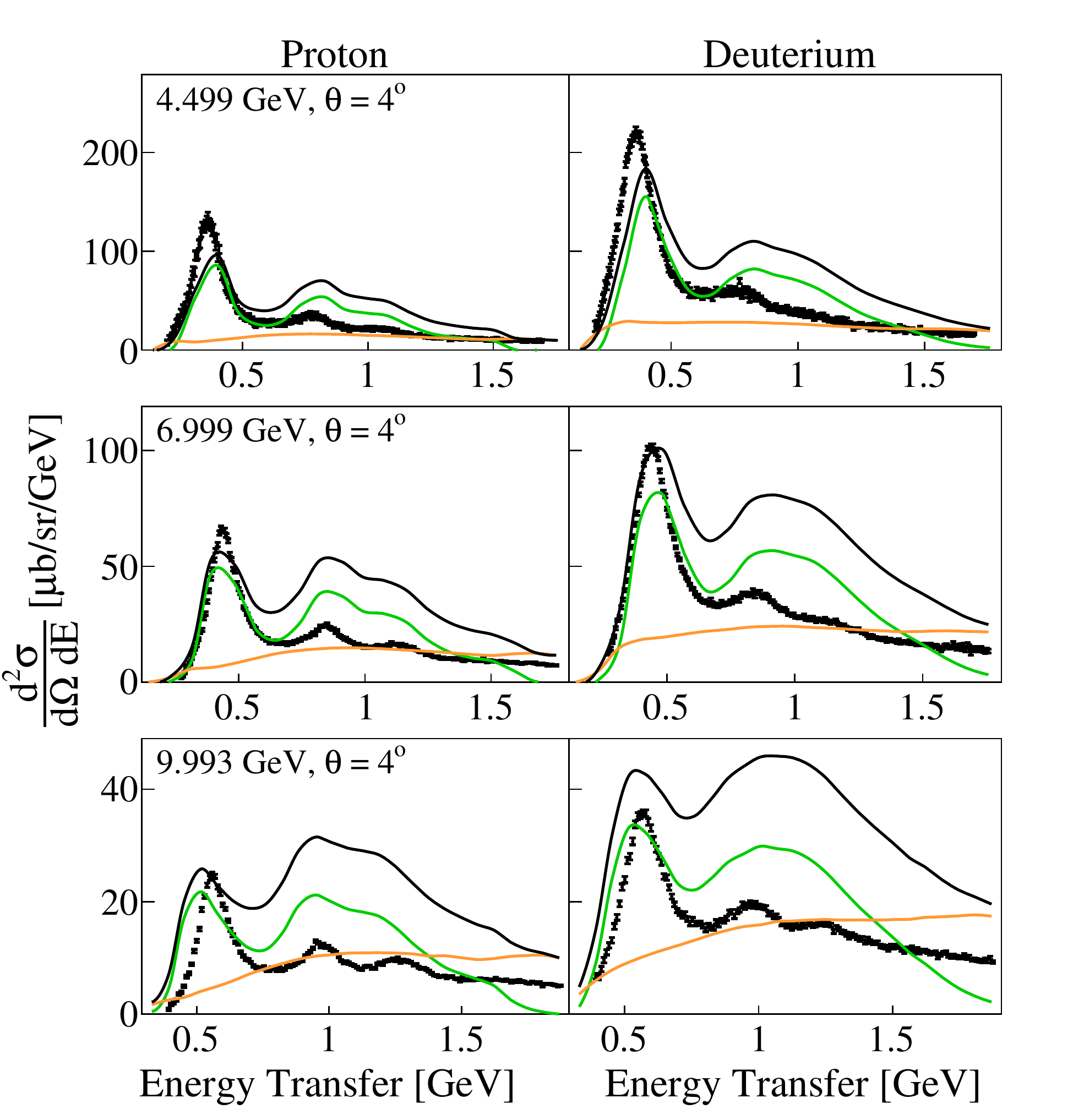}
\caption{Comparison of inclusive proton (left) and deuterium (right)
  $(e,e')$ scattering cross sections for data and for GENIE using
  G2018. (top) $E_0=4.499$ GeV and $\theta_e = 4^\circ$, (middle)
  $E_0=6.699$ GeV and $\theta_e = 4^\circ$, and (bottom) $E_0=9.993$
  GeV and
  $\theta_e = 4^\circ$~\cite{PhysRevD.12.1884}.
  Black points show the data, solid black lines show the total GENIE
  prediction, colored lines show the contribution of the different
  reaction mechanisms: (green)
RES and (orange) DIS.  The first peak at lowest energy transfer is the
$\Delta(1232)$ resonance.}
\label{fig:HydrogenSLAC}
\end{figure}

Fig.~\ref{fig:Ar} shows the inclusive Ar$(e,e')$ cross sections for
$E_0=2.222$ GeV and $\theta_e=15.54^\circ$~\cite{PhysRevC.99.054608}
compared to the G2018 and GSuSAv2 models.  The GSuSAv2 model
reproduces the data very well in the QE-peak region and the G2018
reproduces the data moderately well.  The GSuSAv2 MEC model describes
the dip region much better than the G2018 model.  Again, there is
significant disagreement with the RES and DIS models at larger energy
transfers.

The quality of the agreement between data and GENIE depends
more on the beam energy and angle than on the target mass (from C to
Fe).  There is a possible momentum-transfer dependent shift in the
location of the SuSAV2 QE peak in Fe due to the extrapolation (via
scaling) from C to Fe.  

The GSuSAv2 QE model generally describes the data as well as or better
than the G2018 model.  
The GSuSAv2 MEC model appears to be significantly superior to the
empirical MEC model, especially at $Q^2 < 0.5$ GeV$^2$ or at smaller
scattering angles.  The empirical MEC contribution is often
much smaller than needed to explain the ``dip'' region cross section.
However, as an empirical model, it can be tuned to better describe the data.

$e$-GENIE dramatically overpredicts the large-energy transfer data at
higher momentum transfers ($Q^2>0.5$ GeV$^2$), indicating issues with
the RES (Berger-Sehgal) and DIS (Bodek and Yang) models used.  

This discrepancy at larger momentum and energy transfers is due to the
elementary electron-nucleon cross section in the resonance and DIS
regions, rather than to the nuclear models, since $e$-GENIE also
significantly overpredicts the proton and deuteron cross sections, especially above
the $\Delta$ peak (see
Figs.~\ref{fig:Hydrogen} and \ref{fig:HydrogenSLAC}).  This
shows that tuning the RES and DIS models to neutrino data is not
sufficient to constrain the vector part of the cross section.

\begin{figure} [!htbp]
\centering
\includegraphics[width=0.23\textwidth]{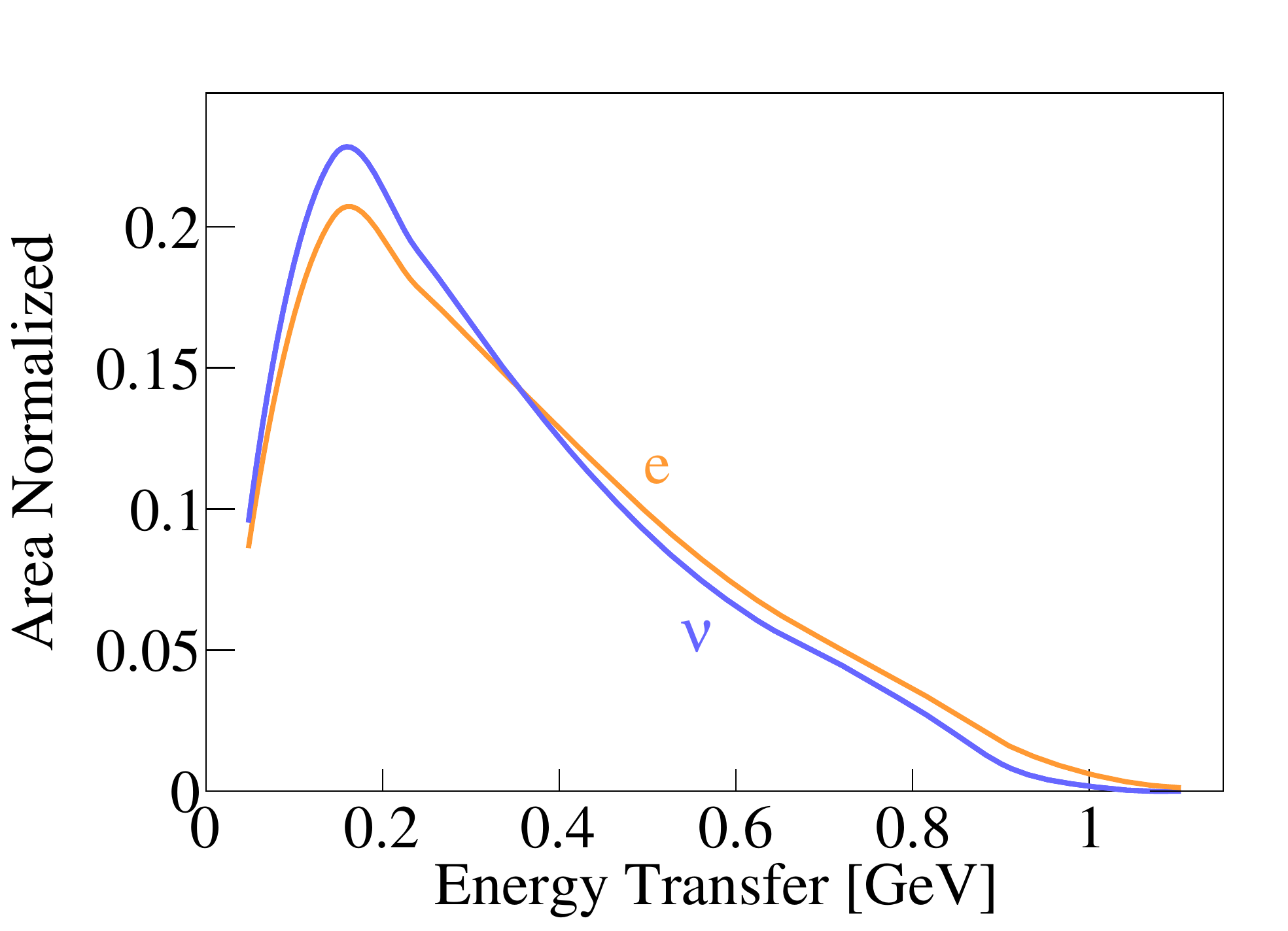}
\includegraphics[width=0.23\textwidth]{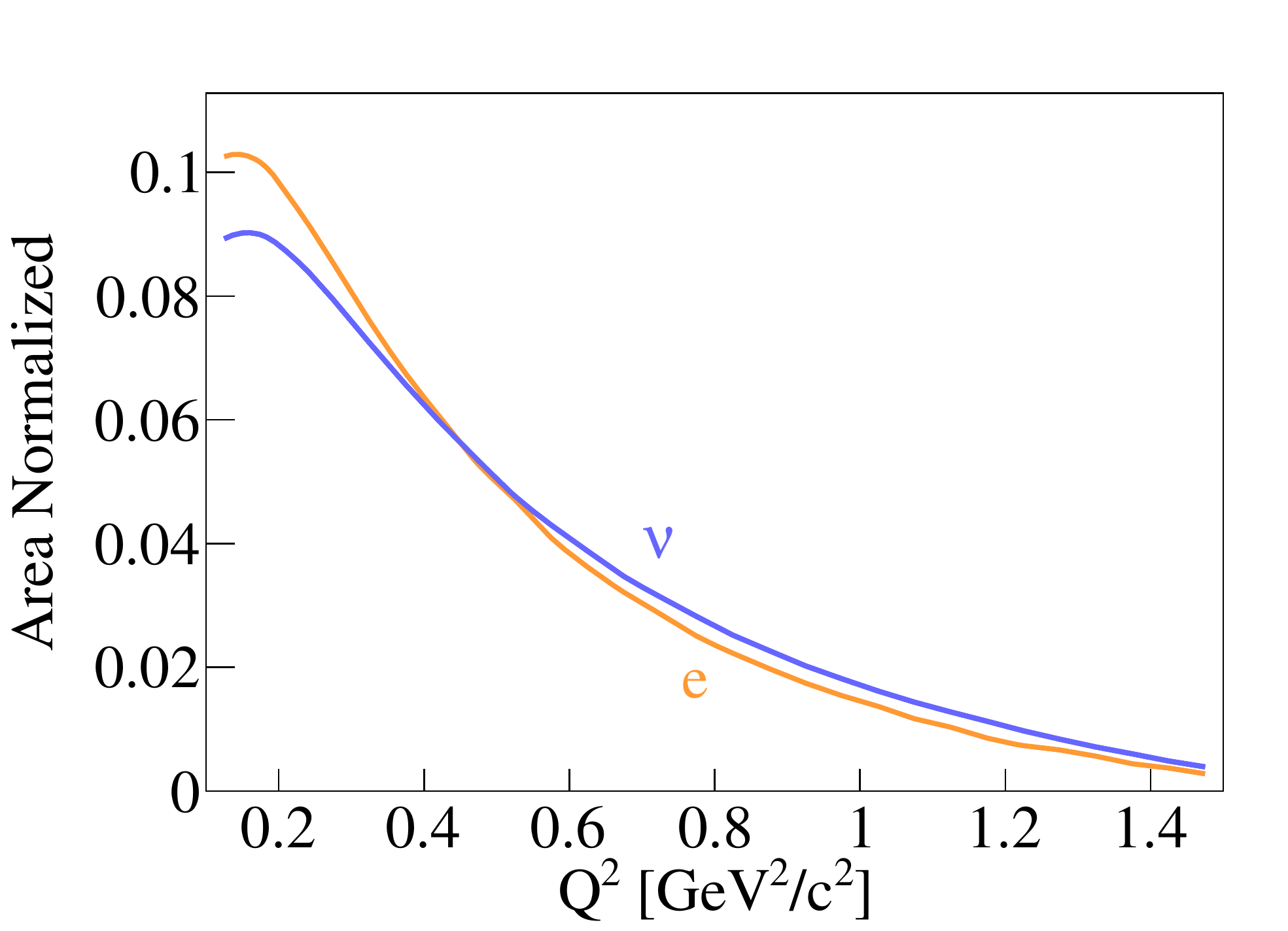}
\caption{\label{e_nu_similarities} Comparison of semi-exclusive 1.16
  GeV lepton-carbon scattering for $Q^2\ge 0.1$ GeV$^2$.  The number
  of generated events is plotted versus energy transfer (left) and
  4-momuntm transfer squared (right) for events with exactly one
  proton with $P_{p} \geq$ 300 MeV/c, no charged pions with
  $P_{\pi} \geq$ 70 MeV/c and no neutral pions or photons of any
  momentum for $e$-GENIE electrons (orange) and GENIE CC $\nu_{\mu}$
  (blue). The electron events have been weighted by $Q^{4}$.  Both
  curves are area normalized.}
\end{figure}

\begin{figure} [htb]
\centering
\includegraphics[width=0.48\textwidth]{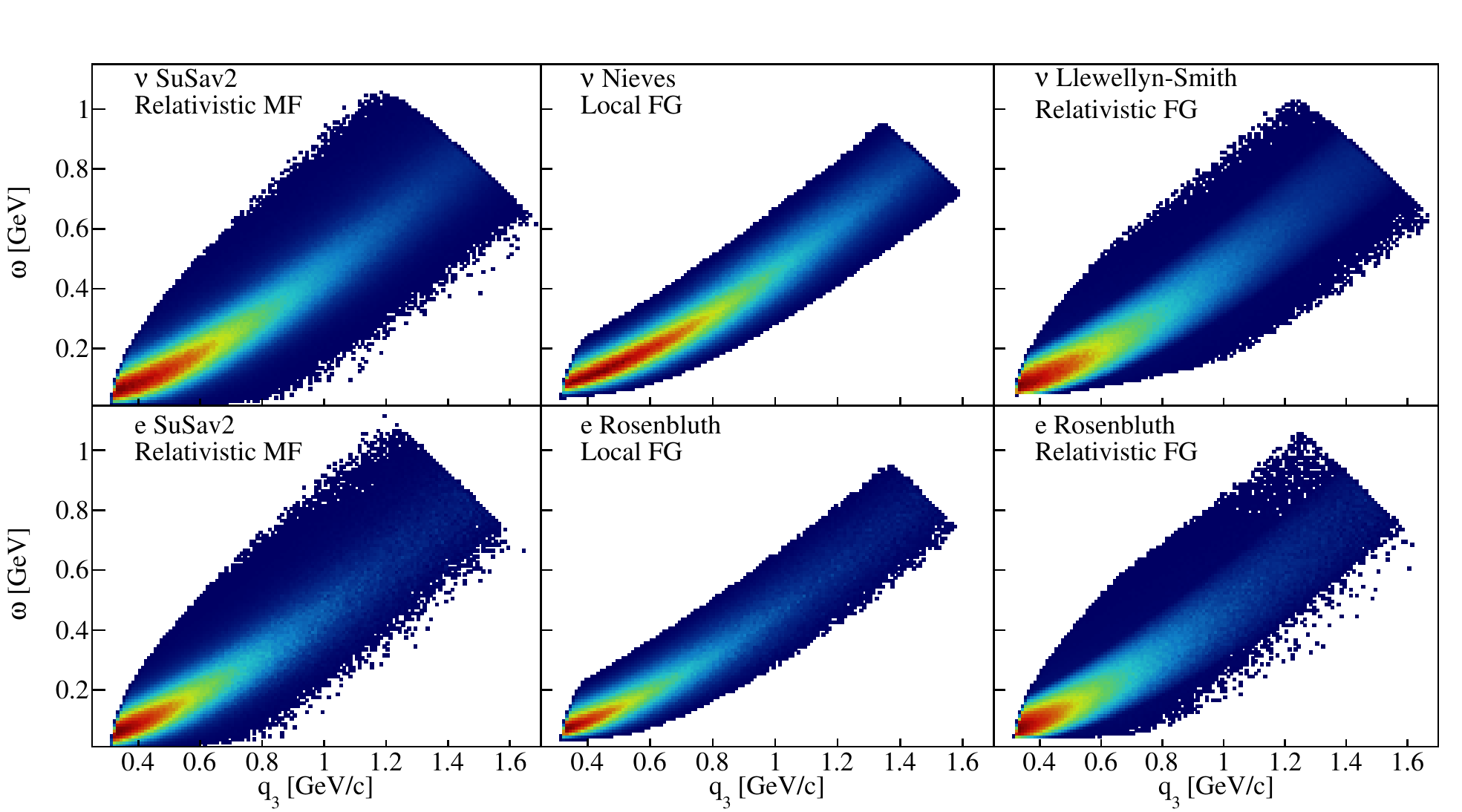}
\caption{\label{QE_Nuclear_Models} Number of simulated events for QE
  scattering on $^{12}$C at 1.161 GeV with $Q^{2} \geq$ 0.1 shown as a
  function of the energy transfer $\omega$ and the momentum transfer
  $q_{3}=\vert\vec q\thinspace\vert$ for all the available nuclear models in GENIE for neutrinos
  (top) and for electrons (bottom). (left) the GSuSAv2
  model which uses a Relativistic Mean Field momentum distribution,
  (middle) the Nieves or Rosenbluth cross section with the Local Fermi
  Gas momentum distribution, and (right) the Llewellyn-Smith or
  Rosenbluth cross section with the Relativistic Fermi Gas momentum
  distribution. The electron events have been
  weighted by $Q^{4}$.}
\end{figure}


\subsection{Implications For Neutrinos}

Electron-scattering data can be a very effective tool for
testing neutrino event generators due to the similarity between the
interactions.   Fig.~\ref{e_nu_similarities}
shows the remarkably similar cross-section shapes for electron-nucleus and
neutrino-nucleus scattering for semi-exclusive 1.16 GeV lepton-carbon
scattering  with exactly one proton with $Q^2\geq 0.1$~GeV$^2$ and $P_{p} 
\geq$
300 MeV/c, no charged pions with $P_{\pi} \geq$ 70 MeV/c and no
neutral pions or photons of any momenta.  This corresponds
approximately to  the Jefferson Lab CLAS detector thresholds.  When
comparing electron and neutrino distributions, the
electron events are each weighted by
$Q^4$ to reflect the difference in the electron and neutrino
elementary interactions.

Exploiting these similarities within the same code is invaluable for minimizing 
the systematic uncertainties of future high-precision neutrino-oscillation
experiments. Oscillation analysis uncertainties exceeding 1\% for
signal and 5\% for backgrounds may substantially degrade the
experimental sensitivity to CP violation and mass hierarchy~\cite{DUNE}. Such
uncertainties include uncertainties in the $\nu$-nucleus interaction.
These uncertainties are typified by the choices of the nuclear model
available in GENIE.

\begin{figure} [t]
\centering
\includegraphics[width=0.48\textwidth]{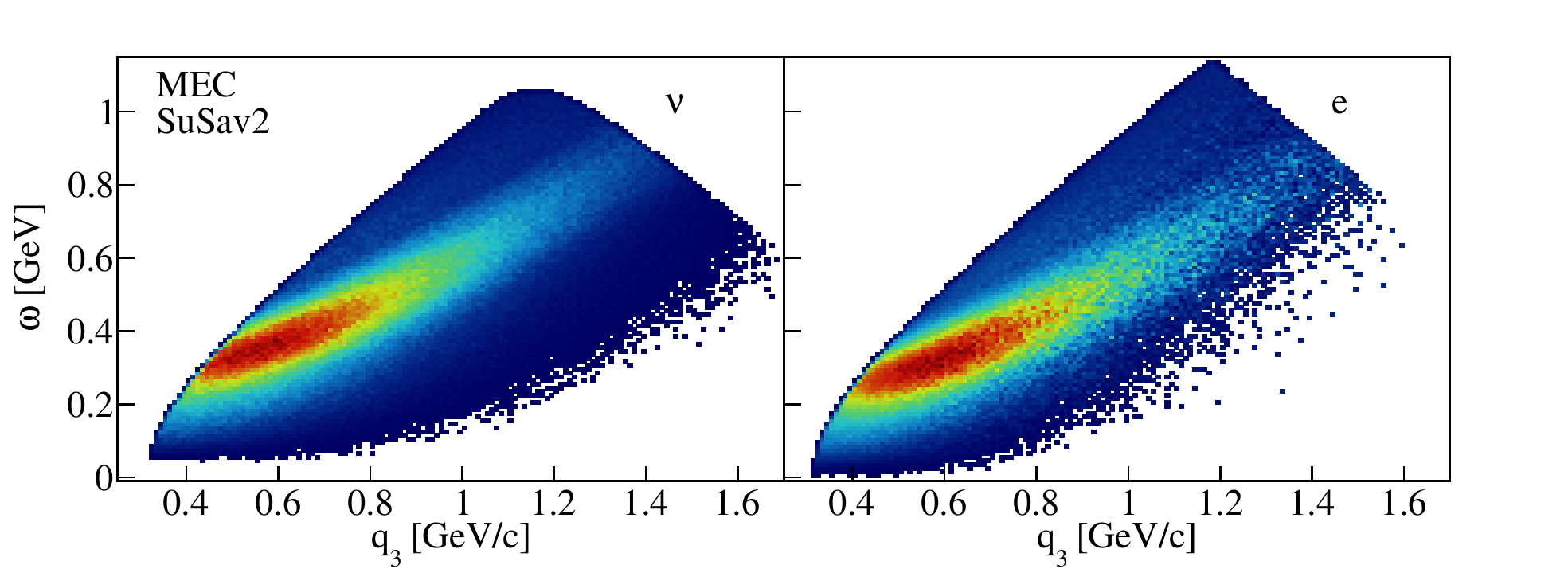}
\caption{\label{MEC_Q0_Q3} Number of simulated events as a function of
  the energy transfer $\omega$ and of the momentum transfer
  $q_{3}=\vert\vec q\thinspace\vert$ for
  neutrinos (left) and for electrons (right) using GSuSav2 for MEC
  interactions. The electron events have been scaled by $Q^{4}$ and
  all the samples have been generated with $Q^{2} \geq$ 0.1.  }
\end{figure}

Fig.~\ref{QE_Nuclear_Models} shows that there is a larger difference
among QE scattering models than there is between QE electron and
neutrino scattering using the same nuclear model.  All six panels show
a ``ridge'', a maximum in the cross section as a function of energy
transfer and momentum transfer.  The length of the ridge (the decrease
in intensity as the energy and momentum transfers increase) reflects
the momentum transfer dependence of the nucleon form factors used in
the cross section model. The width of the distribution perpendicular
to the ridge reflects the width of the nuclear momentum distribution.
The Local Fermi Gas model has a much narrower momentum distribution
than either the Relativistic Mean Field or the Relativistic Fermi Gas
models.  The Nieves cross section decreases more slowly with momentum
transfer than the others.  For GSuSAv2, the electron cross section appears to
decrease slightly faster with momentum transfer than the neutrino
cross section, possibly reflecting differences in the axial and vector
nucleon form factors.  We compared the SuSAv2 and Rosenbluth/LFG
models to electron scattering
cross sections in the previous sections of this paper. 

Similarly,
Fig.~\ref{MEC_Q0_Q3} shows that the distribution of MEC events is very
similar for electrons and for neutrinos within the same model.  Thus,
measurements of electron scattering will be able to significantly
constrain models of neutrino scattering.

Our ability to use the GENIE code to transfer knowledge gained from electron 
scattering depends critically on the implementation of its components.
Because of its modular design, all reaction models in GENIE use the
same nuclear model (e.g., RFG or LFG).
Although the electron scattering capability was added after the initial code
release, many of the reaction models used electron scattering data to
construct  the vector components of neutrino
interactions.  This was true for the resonance~\cite{Rein:1980wg,Berger:2007rq} and the DIS~\cite{Bodek2003} interactions.
The difference between vector neutrino and electron scattering is an overall 
factor  (see Eq.~\ref{eq:factor}) 
and an appropriate change in form factors.

Both QE and MEC  models use the same
vector form factors for neutrino and for electron scattering.  QE models can use
nucleon form factors from electron scattering~\cite{Bradford:2006yz},
but MEC models must 
calculate the form factors.  

The GENIE nonresonant meson-production cross section (referred to as
``DIS'') comes from the Bodek-Yang model ~\cite{Bodek2003} for the full cross section
which extends to $\pi$N threshold.  The cross section is scaled in 
the resonance region so that it agrees with $\nu D$ data~\cite{TenaVidal2018}.
Since a single factor is used to fit the model to the neutrino data,
the high-quality $ep$ and $eD$ data will be poorly described.  While
the total neutrino cross section and some of the hadronic content of
the final state are loosely constrained by the $\nu$D data, the vector
component of the models is poorly constrained.

The QE models describe the data reasonably well in the low
energy-transfer region.  Similarly, the
largest energy-transfer
portions of Figs.~\ref{fig:C3} and \ref{fig:Hydrogen} show a reasonable
agreement between GENIE and data.  However, at intermediate energy
transfer, the resonance region modeling disagrees
with the data for both nuclear and nucleon targets (as in Ref.~\cite{Ankowski:2020qbe}).  
This is due to
the use of resonance form factors that are not up-to-date (RES) and the way
the nonresonant contribution was modeled.  

Improvements are in progress but are not simple and therefore not
available at this time.  A possible short-term
fix would be to include the $ep$ and $eD$ inclusive
electron-scattering models of Bosted and
Christy~\cite{Christy:2007ve,Bosted:2007xd}.  Alternatively, the
vector resonant form factors could be updated using electroproduction
data from Jefferson Lab and elsewhere. A fit to that data is
available~\cite{maid} and partially implemented in GENIE, but it does
not include non-resonant scattering.  A more comprehensive solution
would be to use the recent DCC model~\cite{Kamano:2016bgm,Nakamura:2013zaa} to
simultaneously describe both resonant and non-resonant scattering of
both electrons and neutrinos.


\section{Summary}
We significantly improved and updated the electron version of GENIE,
the popular neutrino-nucleus event generator.  We also added partial
radiative corrections for electron scattering.  Improvements came from
bug fixes and extensions to the QE, 2p2h, and $\Delta$ excitation
models for G2018 and an addition of the SuSav2 model for QE and MEC.
The RES and DIS models are almost identical to past
implementations~\cite{Genie2010} with the main change coming
from a retune to the $\nu$D data~\cite{TenaVidal2018}.

We compared two different GENIE model sets to inclusive
electron-scattering data for a wide range of targets, beam energies
and scattering angles.  The G2018 and GSuSAv2 model sets differ in
their description of QE and MEC scattering.  The SuSAv2 model
generally describes the data at the QE peak as well as or better than
the G2018 model.  The SuSAv2 model set describes the dip region in
most of the data sets much better than the G2018, especially at lower
momentum transfer or smaller electron scattering angle.

At the highest momentum transfers, $e$-GENIE dramatically overpredicts
the data at large energy transfer, indicating significant problems
with the momentum-transfer dependence of the RES and DIS models used.
This discrepancy at larger momentum transfer is due to discrepancies
in the electron-proton and electron-deuteron cross section rather than
to the nuclear models.  This conclusion is similar to that of Ref.~\cite{Ankowski:2020qbe},
but we explore the difference in more detail.  Tuning the RES and DIS models to
neutrino data alone is not sufficient.  Including electron scattering
data will allow tuning GENIE to describe the vector part of the RES
and DIS interaction more precisely.

We found that the shapes of the scattered-lepton energy and momentum
transfer distributions are remarkably similar for electrons and for
neutrinos, when the electron events are each weighted by $Q^4$ to
reflect the difference in the elementary lepton interactions.  In
addition, the differences among QE interaction models are
significantly larger than the difference between electron-nucleus and
neutrino-nucleus scattering. This validates our use of
electron-nucleus scattering data to constrain neutrino-nucleus
scattering models.


The long term goal is to rigorously test the vector current part of the
lepton-nucleus interaction and to use that information to improve
modeling of neutrino-nucleus interactions.  More extensive and
more exclusive electron-scattering data sets are becoming available and 
will be used in the future.  Simultaneously, improvements in GENIE are
ongoing.  The QE and MEC models have been significantly improved with
the theoretically-inspired SuSAv2 models.  Similar improvements are
needed in the RES and DIS modeling.
The combination of new, high-precision, electron-scattering data and
modern interaction models in GENIE should significantly decrease the
systematic uncertainties of future neutrino oscillation experiments.

\textbf{Acknowledgements:} The authors thank the US Department of Energy for
support under contracts DE-FG02-96ER40960, DE-SC0007914, and DE-SC0020240.
G.D.M. acknowledges support from CEA, CNRS/IN2P3 and P2IO, France; by
the European Union's Horizon 2020 research and innovation programme
under the Marie Sklodowska-Curie grant agreement No. 839481; by the
Spanish Ministerio de Economia y Competitividad and ERDF (European
Regional Development Fund) under contract FIS2017-88410-P, and by the
Junta de Andalucia (grants No. FQM160, SOMM17/6105/UGR).
A.P. acknowledges support from the Visiting Scholars Award Program of
the Universities Research Association.


\bibliography{e4nu}

\end{document}